\documentclass[preprint, 12pt]{elsart}

\usepackage{graphicx}

\usepackage{amssymb}
\usepackage{amsmath}
\usepackage{dsfont}
\usepackage{amsfonts}
\usepackage{cancel}
\usepackage{color}
\usepackage{epstopdf}
\usepackage{threeparttable}

\journal{Nuclear Physics A}

\begin{document}

\begin{frontmatter}



\title{Interaction of antiprotons with nuclei}


\author{Jaroslava Hrt\'{a}nkov\'{a} and Ji\v{r}\'{\i} Mare\v{s}}

\address{Nuclear Physics Institute, 250 68 \v{R}e\v{z}, Czech Republic}

\begin{abstract}
We performed fully self-consistent calculations of $\bar{p}$-nuclear bound states
using a complex ${\bar p}$-nucleus potential accounting for $\bar{p}$-atom data.
While the real part of the potential is constructed within the relativistic mean-field
(RMF) model, the $\bar{p}$ annihilation in the nuclear medium is described by a phenomenological optical potential.
We confirm large polarization effects of the nuclear core caused by the presence of the antiproton.
The ${\bar p}$ annihilation is treated dynamically, taking into account explicitly the reduced phase space for
 annihilation from deeply bound states as well as the compressed nuclear density due to the antiproton.
The energy available for the products of ${\bar p}$ annihilation in the nuclear medium is evaluated self-consistently,
considering the additional energy shift due to transformation from the ${\bar p}N$ system to ${\bar p}$-nucleus system.
Corresponding $\bar{p}$ widths in the medium are significantly suppressed, however, they still remain considerable
for the $\bar{p}$ potential consistent with experimental data.
\end{abstract}

\begin{keyword}
antiproton-nucleus interaction \sep antiproton annihilation \sep antiproton nuclear bound states
\end{keyword}

\end{frontmatter}

\section{Introduction}
\label{intro}
The study of the interaction of antiprotons with nuclei is a source of valuable information about the
behavior of antiproton in nuclear matter, the in-medium $\bar{p}N$ interactions, as well as nuclear dynamics.
Experiments aiming at exploring the $\bar{p}$-nucleon interaction has been performed since the discovery of
the antiproton in 1955 \cite{objav antiprotonu}. The antiproton-proton annihilation was studied at the Brookhaven National Laboratory (BNL) and CERN in the 1960`s~\cite{Bubble chamber} (see also \cite{KlemptAnnih} and references therein).

Theoretical considerations about the $\bar{p}$-nucleus interaction are based on symmetry between $NN$ and $\bar{N}N$ potentials. In the framework of a meson exchange
model, the real part of an $\bar{N}N$ potential constructed using
the G-parity transformation is strongly attractive \cite{machlaidt88}, which led to
conjectures about deeply bound ${\bar p}$ states in nuclei \cite{burvenich, Wong, Baltz}.
The possibility of existence of antiproton-nucleon or antiproton-nucleus quasi-bound states was studied in
experiments at the LEAR facility at CERN \cite{antiNN interaction}. The $\bar{p}$ elastic and inelastic scattering off nuclei and proton knock-out reactions were analyzed in order to extract information about the $\bar{p}$-nucleus potential.
The measurements of the differential cross-section for the $\bar{p}$ elastic scattering off $^{12}$C at 46.8 MeV favor a
shallow attractive Re$V_{\text{opt}}$ with the depth $\leq70$~MeV and an absorptive part Im$V_{\text{opt}}\geq 2$Re$V_{\text{opt}}$~\cite{antiNN interaction}.
On the other hand, $\bar{p}$ production in proton-nucleus and nucleus-nucleus collisions is well described by Re$V_{\text{opt}}\sim -$~(100 - 200)~MeV~\cite{Cassing}.
Despite considerable efforts, no convincing evidence for existence of $\bar{p}$N or $\bar{p}$-nucleus bound states has been found \cite{lw05, xhm15}.

Unique information about the $\bar{p}$--nucleus optical potential near threshold has been provided by
analyses of strong interaction energy shifts and widths of ${\bar p}$-atomic levels \cite{batty, friedman, mares}.
Global fits of 107 data points of X-ray and radiochemical data led to the ${\bar p}$ potential with an
attractive real part about 110 MeV deep and an absorptive imaginary part about 160 MeV deep when extrapolated
into the nuclear interior \cite{mares}.
However, the ${\bar p}$-atom data probe reliably the $\bar{p}$--nucleus potential at the far periphery of the
nucleus and model dependent extrapolations to the nuclear interior are a source of large uncertainties.
Very recently Friedman et al ~\cite{Paris1} applied $\bar{N}N$ scattering amplitudes of the latest version of the Paris $\bar{N}N$ potential \cite{Paris2}
to construct the $\bar{p}$-nucleus optical potential and demonstrated the importance of P-wave amplitudes
 to account for the $\bar{p}$-atom data.

The $\bar{p}$-nucleus interaction has attracted renewed interest in recent years at the prospect of future
experiments with ${\bar p}$ beams at the FAIR facility at GSI \cite{FAIR}.
The ${\bar p}$-nuclear bound states and the possibility of their formation have been studied in Refs.~\cite{Mishustin, larionov, larionov2, lari mish pschenichov, lari mish satarov, lenske}
within the relativistic mean-field approach \cite{Walecka, reinhard} by employing the G-parity transformation of nucleon-meson
coupling constants.
A scaling factor $\xi$ was introduced to vary the depth of the $\bar{p}$-nucleus potential \cite{Mishustin, larionov, larionov2, lari mish pschenichov, lari mish satarov}. This scaling factor which represents departure from the G-parity symmetry can be then fitted to yield the $\bar{p}$ potential consistent with available experimental data. The calculations predicted strong binding of the antiproton inside a nucleus and large compression of the nuclear core induced by the
presence of $\bar{p}$. 
The $\bar{p}$ annihilation in the nuclear medium was studied as well~\cite{Mishustin}. Partial widths were evaluated with the help
of vacuum annihilation cross sections for considered annihilation channels and the phase space suppression for ${\bar p}$ annihilation from deeply bound states
was taken account. The lifetime of $\bar{p}$ in a nucleus was estimated to be in the range of 2 - 20~fm/c.

In Refs. \cite{lari mish pschenichov, lari mish satarov} the Giessen Boltzmann-Uehling-Uhlenbeck (GiBUU) transport model \cite{gibuu} was
applied to ${\bar p}$-nucleus interactions in a wide range of ${\bar p}$-beam momenta.
The GiBUU model was used to fit the KEK data \cite{KEK} on $\bar{p}$ absorption cross sections at $p_{\rm lab}=470\;\text{-}\; 880$~MeV/c to fix the value of the
scaling factor $\xi=0.22$, which corresponds to  Re$V_{\text{opt}} \simeq 150$~MeV deep at normal nuclear density.
Dynamical response of selected nuclei to the incident antiproton together with the probability that
the antiproton reaches the dense nuclear environment before it annihilates was examined.
The time required for the nuclear compression was found to be within the range of the ${\bar p}$ lifetime calculated in Ref. \cite{Mishustin}.

Recently, Gaitanos et al.~\cite{NLD1, NLD2} developed a non-linear derivative (NLD) model which accounts for momentum dependence of the nuclear mean fields, which
is missing in standard RMF models.
This momentum dependence reduces the G-parity motivated $\bar{p}$ optical potential and yields its depth in agreement with available
experimental data. It was demonstrated that the RMF approach with antiproton-meson couplings scaled by a factor $\xi=0.2\; \text{-}\; 0.3$ can reproduce
the NLD results in average~\cite{NLD3}.

In this work, we performed fully self-consistent calculations of ${\bar p}$ nuclear bound states using a complex
${\bar p}$-nucleus potential consistent with ${\bar p}$-atom data, aiming at analyzing in detail
various effects which could have impact on calculated ${\bar p}$-nuclear characteristics.
In particular, we explored dynamical response of the nuclear core to the presence of the deeply bound antiproton.
In view of appreciable densities in the interior of ${\bar p}$ nuclei, it is desirable to check how reliable are
the underlying RMF models in such highly dense nuclear matter. We therefore applied in our calculations various
RMF models which yield different compressibilities of nuclear matter, including the TW99 model with density-dependent
couplings \cite{TypelWolter}, and compared their predictions.
Annihilation widths  in the nuclear medium depend strongly on the energy available for the
decay products of the deeply bound antiproton, as well as the density of the surrounding nuclear medium. It is thus
imperative to perform fully dynamical calculations of ${\bar p}$-nuclear states using a complex ${\bar p}$ potential which incorporates main features of the ${\bar p}$-nucleus interaction, while taking into account self-consistently the additional energy shift corresponding to the transformation from the 2 c.m. ${\bar p}N$ annihilation to ${\bar p}N$ annihilation in a
nucleus. The procedure for self-consistent handling the sub-threshold energy dependence was recently applied in calculations of
kaonic atoms, and kaonic and $\eta$ nuclear states using chirally motivated ${\bar K}N$ amplitudes \cite{s, kaony, kaonic atoms1, kaonic atoms2, etaS1, etaS2}.

The paper is organized as follows. In Section 2, we briefly describe the applied RMF model for calculating ${\bar p}$ nuclear states, discuss the underlying
${\bar p}$-nucleus interaction and ${\bar p}$ absorption in the nuclear medium including self-consistent schemes for evaluating the energy
$\sqrt{s}$ which enters phase space suppression factors.
In Section 3, we present selected results of our calculations of ${\bar p}$ quasi-bound
states in various nuclei across the periodic table in order to demonstrate dynamical effects in the nuclear core caused by the antiproton, model dependence of the calculations, and the role of various factors that determine ${\bar p}$ widths in the nuclear medium.
Conclusions are summarized in Section 4.

\vspace{-10pt}
\section{Model}
\label{model}
The interaction of an antiproton with a nucleus is studied within the relativistic mean-field model \cite{Walecka, reinhard}. In this model, the interaction among (anti)nucleons is mediated by the exchange of the scalar ($\sigma$) and vector ($\omega_{\mu}$, $\vec{\rho}_\mu$) meson fields, and the massless photon field $A_{\mu}$. In order to incorporate the $\bar{p}$ into the model we extended the standard Lagrangian density for nucleonic sector by the Lagrangian density which describes the antiproton interaction with the nuclear medium:
\begin{equation} \label{LagDens}
\begin{split}
\mathcal L=& \sum_{j=N,\bar{p}} {\bar{\psi_j}} [i\gamma^\mu \partial_\mu \!-\! m_j \!-\! g_{\sigma j}{\sigma} \!-\! g_{\omega j } \gamma_\mu{\omega}^\mu \!-\! g_{\rho j}\gamma_\mu \vec{\tau}\cdot{\vec{\rho}}^\mu \!-\! e\gamma_\mu \frac{1}{2}(1+\tau_3) {A}^\mu]{\psi_j}\\ &+ \frac{1}{2}\left(\partial_\mu{\sigma} \partial^\mu {\sigma} -m_\sigma^2 {\sigma}^2\right) - \frac{1}{2}(\frac{1}{2} {\Omega}_{\mu \nu} {\Omega}^{\mu \nu} - m_{\omega}^2 {\omega}^\mu {\omega}_\mu ) \\ &- \frac{1}{2}(\frac{1}{2} {\vec{R}}_{\mu \nu}\cdot{\vec{R}}^{\mu \nu} - m_\rho^2{\vec{\rho}}_\mu
\cdot{\vec{\rho}}^\mu) - \frac{1}{4} {F}_{\mu \nu} {F}^{\mu\nu} \\ & - \frac{1}{3} g_2 {\sigma}^3 - \frac{1}{4} g_3 {\sigma}^4 + \frac{1}{4}d({\omega}^\mu {\omega}_\mu)^2~,
\end{split}
\end{equation}
where $m_j$ denotes the mass of the (anti)nucleon; $m_\sigma$, $m_\omega$, $m_\rho$ are the masses of the considered meson fields; $g_{\sigma j}$, $g_{\omega j}$, $g_{\rho j}$ and $e$ are the (anti)nucleon couplings to corresponding fields --- $g_2, g_3$ and $d$ represent the strengths of the $\sigma$  and $\omega$ field self-interactions. The field tensor fulfills ${F}_{\mu \nu}=\partial_\mu {F}_\nu - \partial_\nu {F}_\mu$, and correspondingly for the ${\Omega}_{\mu \nu}$ and ${\vec{R}}_{\mu \nu}$.

The equations of motion are derived using the variational principle employing the mean-field and no-sea approximations. Furthermore, we are dealing with stationary states and spherically symmetric nuclei. We assume that single particle states do not mix isospin, i.\ e., only the neutral component of the isovector $\rho$-meson field is considered. The Dirac equations for nucleons and antiproton then read:
\begin{equation} \label{DiracEq}
[-i\vec{\alpha}\vec{\nabla} +\beta(m_j + S_j) + V_j]\psi_j^{\alpha}=\epsilon_j^{\alpha} \psi_j^{\alpha},
\quad j=N,\bar{p}~,
\end{equation}
where
\begin{equation}
S_j=g_{\sigma j}\sigma, \quad V_j=g_{\omega j} \omega_0 + g_{\rho j}\rho_0 \tau_3 + e_j \frac{1+\tau_3}{2}A_0
\end{equation}
are the scalar and vector potentials and $\alpha$ denotes single particle states.
The equations of motion for the boson fields acquire additional source terms due to the presence of $\bar{p}$:
\begin{equation}
\begin{split} \label{meson eq}
(-\triangle + m_\sigma^2+ g_2\sigma + g_3\sigma^2)\sigma&=- g_{\sigma N} \rho_{SN}-g_{\sigma \bar{p}}
\rho_{S \bar{p}}~, \\
(-\triangle + m_\omega^2 +d\omega^2_0)\omega_0&= g_{\omega N}\rho_{VN} +g_{\omega \bar{p}} \rho_{V\bar{p}}~, \\
(-\triangle + m_\rho^2)\rho_0&= g_{\rho N}\rho_{IN} +g_{\rho \bar{p}}\rho_{I \bar{p}}~, \\
-\triangle A_0&= e_N \rho_{QN}+e_{\bar{p}}\rho_{Q\bar{p}}~,
\end{split}
\end{equation}
where $\rho_{\text{S}j}, \rho_{\text{V}j}, \rho_{\text{I}j}$ and $\rho_{\text{Q}j}$  are the scalar,
vector, isovector, and charge densities, respectively. The coupled system of the equations of
motion Eq.~\eqref{DiracEq} and Eq.~\eqref{meson eq} is solved fully self-consistently by iterative
procedure.

The nucleon-meson coupling constants and meson masses were adopted from the nonlinear RMF model TM1 for heavy nuclei and TM2 for light nuclei \cite{Toki}. These two RMF parametrizations proved successful in
the description of ground state characteristics of ordinary nuclei in the corresponding mass regions, however,
it is not guaranteed that they will provide consistent account of the properties of $\bar{p}$ nuclei (e.g. $\bar{p}$ and total binding energies),
particularly their $A$ dependence. Moreover, since TM1 and TM2 yield quite different compressibilities of nuclear
matter, they could predict different size of the nuclear core modifications due to the antiproton.
We thus performed calculations of selected $\bar{p}$ nuclei using the RMF NL-SH parametrization \cite{nlsh} as well,
and studied model dependence of our results.

The antiproton placed in a nucleus causes strong polarization effects resulting in the high central density of the
nuclear core, reaching up to 4 times the nuclear matter density. The application of standard RMF models for the description of nuclear matter
at such densities has to be considered as extrapolation. Therefore, we employed also the density-dependent RMF model \cite{TypelWolter} which is more suitable for the
 description of dense nuclear matter. In the density-dependent model, the nucleon-meson couplings are a function of the nucleon density $\rho_{\text VN}$
\begin{equation}
g_{iN}(\rho_{\text{V}N})=g_{iN}(\rho_{0})f_i(x)~, \quad i=\sigma, \omega~,
\end{equation}
where
\begin{equation}
f_i(x)=a_i\frac{1+b_i(x+d_i)^2}{1+c_i(x+d_i)^2}~,
\end{equation}
and  $x=\rho_{\text{V}N}\slash\rho_{0}$, where $\rho_0$ represents the saturation density of nuclear matter.
The coupling of the $\rho$ meson has an exponential character
\begin{equation}
g_{\rho N}(\rho_{VN})=g_{\rho N}(\rho_{0})\text{exp}[-a_{\rho}(x-1)]~.
\end{equation}
The parameters $a_i, b_i, c_i, d_i$ and $a_{\rho}$ are fitted to Dirac-Brueckner calculations of nuclear matter and constrained by conditions on the functions $f_i(x)$ \cite{TypelWolter}.
The density dependence of the nucleon-meson couplings leads to an extra term $\Sigma_{\text{R}}$ in the Dirac equation for nucleons
\begin{equation} \label{Dirac DD coupling}
[-i\vec{\alpha}\vec{\nabla} +\beta(m_N + S_N) + V_N+\Sigma_{\text{R}}]\psi_N^{\alpha}=\epsilon_N^{\alpha} \psi_N^{\alpha}~,
\end{equation}
where
\begin{equation}
\Sigma_{\text{R}}=\frac{\partial g_{\omega N}}{\partial \rho_{\text{V}N}}\rho_{\text{V}N}\omega_0 + \frac{\partial g_{\rho N}}{\partial \rho_{\text{V}N}}\rho_{\text{I} N}\rho_0 - \frac{\partial g_{\sigma N}}{\partial \rho_{\text{V}N}}\rho_{\text{S} N}\sigma~.
\end{equation}
The Klein-Gordon equations for the meson fields retain their form as in Eq.~\eqref{meson eq} only the couplings become a
function of density.

The equations of motion in the RMF model are derived on the Hartree level where each nucleon moves in mean fields
created by \emph{all} nucleons bound in the nucleus. Consequently, the nucleon feels in addition a kind of
``attraction'' as well as ``repulsion'' from itself. In ordinary nuclei this self-interaction has only a minor
(1/A) effect.\footnote{It is to be noted that the self-interaction is directly subtracted in the Hartree-Fock formalism.} However, the potential acting on the antiproton in a nucleus is much deeper than the potential acting on nucleons and so the impact of the $\bar{p}$ self-interaction could become pronounced. In order to explore the
role of the $\bar{p}$ self-interaction, we performed calculations where the $\bar{p}$ source terms were omitted in
the Klein-Gordon equations for the boson fields acting on the antiproton, i.e.
\begin{equation}
\begin{split}
(-\triangle + m_M^2)\Phi_{\bar{p}} = g_{M N}\rho_{MN} {\color{red}\xcancel{ +\; g_{M \bar{p}}\rho_{M \bar{p}}}}
\end{split}
\end{equation}
and compared them with the results of regular calculations according to Eq.~\eqref{meson eq}.
The impact of the unphysical $\bar{p}$ self-interaction depends on the depth of the $\bar{p}$ potential.
The deeper is the $\bar{p}$ potential the larger is the role of the $\bar{p}$ self-interaction.
We will demonstrate in the following section that the effect of the $\bar{p}$ self-interaction is negligible for
the $\bar{p}$ potential consistent with available experimental data.

\subsection{$\bar{p}$--nucleus interaction}

On the level of hadron degrees of freedom the strong interaction between nucleons is understood as a
meson exchange process. When going from the $NN$ interaction to $\bar{N}N$ interaction the G-parity
transformation,  which consist of charge conjugation and rotation in isospin space, seems to be a natural link for the
medium and long range part of the interaction which is governed by the meson exchange.
To describe the $\bar{p}$-nucleus interaction we thus make use of the G-parity transformation.
The real part of the $\bar{p}$-nucleus potential is obtained by the transformation of the nucleon-nucleus potential
\begin{equation}
U_{\bar{p}}=\sum_{M} G_{M}U_{M}~,
\end{equation}
where $U_M$ denotes the potential generated by the exchange of the meson $M$ and $G_{M}$ is the G-parity eigenvalue
for the corresponding meson field. When expressed in terms of coupling constants we have
\begin{equation} \label{pbar couplings}
g_{\sigma \bar{p}}=g_{\sigma N}, \quad g_{\omega \bar{p}}=-g_{\omega N}, \quad g_{\rho \bar{p}}=g_{\rho N}~.
\end{equation}
Within the RMF approach the nuclear ground state is well described by an attractive scalar potential $S(0) \simeq -400$~MeV and
a repulsive vector potential $V(0) \simeq 350$~MeV. The central potential acting on a nucleon in a nucleus is then
$S(0) +V(0) \simeq -50$~MeV. Since the vector potential generated by the $\omega$ meson exchange changes its sign under the G-parity transformation, the total $\bar{p}$ potential would be strongly attractive and
$\approx 750$~MeV deep in the nuclear interior.

We should stress that G-parity is surely a valid concept for the long and medium range $\bar{p}$ potential. However,
the $\bar{p}$ annihilation plays a crucial role in the $\bar{p}N$ and $\bar{p}$-nucleus interactions. It has a major
contribution in the short range region and it is not clear to what extent it affects the elastic part of the
interaction. Moreover, various many-body effects could cause deviations from the G-parity values
in the nuclear medium as well \cite{Mishustin}. Therefore G-parity should be regarded as a mere starting point to determine the $\bar{p}$-meson
coupling constants in standard RMF models. It is to be noted that a recent approach \cite{NLD3} which incorporates the momentum dependence of the mean fields yields the $\bar{p}$ potential consistent with in-medium antinucleon phenomenology while retaining the G-parity symmetry.

The form of the $\bar{p}$ potential in the nuclear medium is still quite uncertain, despite
considerable experimental as well as theoretical efforts in the past. Following Refs.~\cite{Mishustin, larionov, larionov2, lari mish pschenichov, lari mish satarov} we introduce a uniform scaling factor $\xi \in \langle0,1\rangle$ for the $\bar{p}$-meson coupling constants:
\begin{equation} \label{reduced couplings}
g_{\sigma \bar{p}}=\xi\, g_{\sigma N}, \quad g_{\omega \bar{p}}=-\xi\, g_{\omega N}, \quad g_{\rho
\bar{p}}=\xi\, g_{\rho N}
\end{equation}
to control the strength of the $\bar{p}$-nucleus interaction. The  experiments with antiprotonic atoms, $\bar{p}$ scattering off nuclei and $\bar{p}$ production in proton-nucleus and nucleus-nucleus collisions
suggest that the real part of the $\bar{p}$ potential should be in the range of $-(100 \; \text{-}\; 300)$~MeV at normal
nuclear density \cite{batty, mares, lari mish pschenichov} which corresponds to $\xi=0.2\; {\text-}\; 0.3$. 

For the real part of the $\bar{p}$-nucleus potential we adopted the value $\xi=0.2$ which provides the $\bar{p}$ potential consistent with $\bar{p}$-atom data. It is to be stressed here that due to sizable modifications of the nuclear core caused by the deeply bound antiproton, the dynamically evaluated $\bar{p}$ potential becomes considerably
deeper than the corresponding potential deduced from the analysis of $\bar{p}$ atoms, as will be demonstrated in Section 4.

\subsubsection{$\bar{p}$ annihilation}
The annihilation of the antiproton in the nuclear medium is an inseparable part of any realistic description of the $\bar{p}$-nucleus interaction. Since the RMF approach does not address directly the $\bar{p}$  absorption in a
nucleus, we adopted the imaginary part of the optical potential in a `$t\rho$' form from optical model phenomenology \cite{mares}:
\begin{equation}
2\mu {\rm Im}V_{\text{opt}}(r)=-4 \pi \left(1+ \frac{\mu}{m_N}\frac{A-1}{A}
\right){\rm Im}b_0 \rho(r)~,
\end{equation}
where $\mu$ is the $\bar{p}$-nucleus reduced mass. While the density $\rho(r)$ was treated as a dynamical quantity evaluated within the RMF model,
the global parameter Im$b_0=1.9$~fm adopted from Ref.~\cite{mares}, was fitted to $\bar{p}$ atom data.
It is to be noted that the value of Im$b_0$ was determined for a finite-range (FR) interaction, where original densities were replaced by 'folded'
densities, while here it was applied to construct a zero-range '$t\rho$' potential.
We checked that the RMF densities in the present work yield r.m.s. radii larger than the unfolded densities used in the $\bar p$ atom analysis and thus
effectively approximate the FR 'folded' densities in Ref.~\cite{mares}.

In our calculations we considered that Im$b_0$ involves annihilation channels with corresponding branching ratios $B_c$ listed in Table \ref{Tab.: annihilation channels}. They are sorted according the number of mesons in final state.
We included only direct decay channels, i.e. only non-resonant contributions and no further decay of produced mesons
were taken into account, as in Ref. \cite{Mishustin}. Moreover, we considered annihilation channels containing kaons.

\begin{table}[hbt]
\begin{center}
\caption{The annihilation channels for $\bar{p}N$ at rest in vacuum. Here, $n_f$ is the number of decay products and
$B_c$ denotes the branching ratio of a particular decay channel$^{\dagger}$.}\vspace{3pt}
\begin{threeparttable}
\begin{tabular}{llc|llc}
$n_f$  &channel     & $B_c$ [\%]           & $n_f$ & channel          & $B_c$ [\%] \\ \hline \hline
2 &2$\pi^0$             & 0.07                 & &$\pi^+ \pi^- \pi^0$  & 1.8   \\
 &$\pi^+\pi^-$         & 0.31                 &   &$\pi_0 K_S K_L$     & 6.7$\cdot 10^{-4}$  \\
 &$\pi^0 \rho^0$       & 1.7                  &   &$\pi^{\pm} K^{\mp} K_S$ & 2.7$\cdot 10^{-3}$ \\
 &$\pi^{\pm}\rho^{\mp}$& 0.9                  &   &$\omega K^+ K^-$        & 2.3$\cdot 10^{-3}$   \\
 &$\pi^0 \omega$       & 0.6                  & 4&4$\pi^0$                & 0.5  \\
 &$\rho^0 \omega$      & 2.3                  &   &$\pi^+ \pi^- 2\pi^0$    & 7.8   \\
 &$\omega \eta$        & 1.5                  &   &2$\pi^+$2$\pi^-$        & 4.2   \\
 &2$\omega$            & 3.0                  &  5& 5$\pi^0$             & 0.5    \\
 &$K^+ K^-$            & 1.0 $\cdot 10^{-3}$  &  & $\pi^+ \pi^- 3\pi^0$    & 20.1  \\
 &$K_S K_L$            & 7.9$\cdot 10^{-4}$   &   &2$\pi^+$2$\pi^- \pi^0$  &10.4   \\
3 &2$\pi^0 \eta$      & 0.7                   &  6&$\pi^+ \pi^-$4$\pi^0$   & 1.9  \\
 &$\pi^+ \pi^- \eta$   & 1.3                   &   &2$\pi^+$2$\pi^-$2$\pi^0$& 13.3  \\
 &2$\pi^0 \omega$      & 2.6                   &  &3$\pi^+$3$\pi^-$        & 2.0  \\
 &$\pi^+ \pi^- \omega$ & 6.6                   &  7& 3$\pi^+$3$\pi^- \pi^0$  & 1.9  \\
 &$\pi^+ \pi^- \rho^0$ & 3.6                   &   & 2$\pi^+$2$\pi^-$3$\pi^0$ & 4.0 \\ \hline \hline
 \end{tabular}
\begin{tablenotes}
\item $^{\dagger}$\footnotesize{The non-strange annihilation channels and their branching ratios are taken from Ref.~\cite{Mishustin} (see also referencies therein). Branching ratios for channels containing kaons are taken from Ref.~\cite{amsler}}
\end{tablenotes}
\end{threeparttable}
\label{Tab.: annihilation channels}
\end{center}
\end{table}

The energy available for ${\bar p}N$  annihilation in vacuum at rest is $\sqrt{s}=m_{\bar{p}}+m_N$. In the nuclear medium,
this energy is reduced due to the binding of the antiproton and nucleon. Consequently, the phase space available
for annihilation products should be substantially suppressed for deeply bound $\bar{p}$, which might lead to a relatively
long living antiproton in the nuclear interior~\cite{Mishustin}.

\begin{figure}[bht]
\begin{center}
\includegraphics[width=0.8\textwidth]{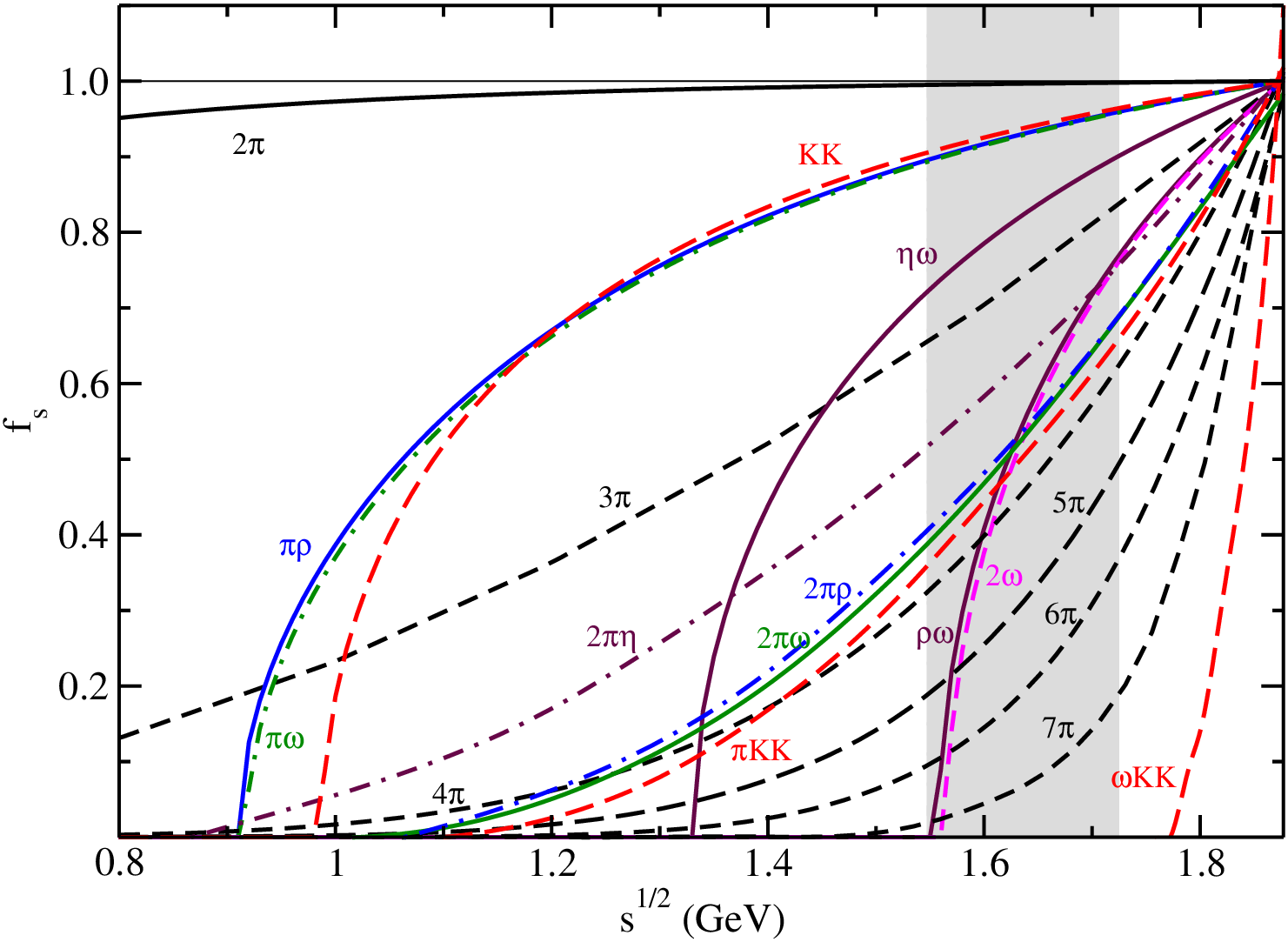}
\caption{\label{Fig.: SF}The phase space suppression factors $f_{\text{s}}$ as a function of the c.m. energy $\sqrt{s}$. The range of $\sqrt{s}$ relevant for ${\bar p}$-nuclear
states is denoted by grey area. }
\end{center}
\end{figure}

We took into account the suppression of phase space by introducing corresponding suppression factors $f_{\text{s}}$.
For the two body decay channels $f_s$ were evaluated with the help of the formula \cite{pdg}:
\begin{equation}
f_{\text{s}}=\frac{M^2}{s}\sqrt{\frac{[s-(m_1+m_2)^2][s-(m_1-m_2)^2]}{[M^2-(m_1+m_2)^2][M^2-(m_1-m_2)^2]}}\Theta(\sqrt{s}-m_1-m_2)~,
\end{equation}
where $m_1$, $m_2$ are the masses of the annihilation products and $M=m_{\bar{p}}+m_N$. For channels containing more than 2 particles in the final state the suppression factors $f_{\text{s}}$ were calculated with the help of the Monte Carlo simulation tool PLUTO~\cite{PLUTO}. To compute the suppression factors for the channels containing more than 4 particles in the final state we expressed the decay products
in terms of two or three effective particles. The $n$-body phase space $\phi_n$ was then decomposed into smaller subspaces according the formula \cite{pdg}
\begin{equation}
\!d\phi_n(P; p_1, \ldots, p_n)\! =\! d\phi_j(q; p_1, \ldots, p_j)\! \times \! d\phi_{n-j+1}(P; q, p_{j+1}, \ldots, p_n) (2\pi)^3 dq^2,
\end{equation}
where $q^2=(\sum_{i=1}^j E_i)^2 - |\sum_{i=1}^j \vec{p}_i|^2$, $P$ is the 4-momentum of the annihilating pair and $p_i$ are the 4-momenta of the annihilation products. The suppression factor was expressed as a ratio of Dalitz plot area for reduced $\sqrt{s}$ and vacuum $\sqrt{s}=2m_N$. The phase space suppression factors for considered annihilation channels are plotted
as a function of the center-of-mass energy $\sqrt{s}$ in Fig.\ref{Fig.: SF}.

The energy available for the annihilation in the medium is given by Mandelstam variable
\begin{equation} \label{s}
s=(E_N + E_{\bar{p}})^2 - (\vec{p}_N + \vec{p}_{\bar{p}})^2~,
\end{equation}
where $E_N=m_N - B_N$, $E_{\bar{p}}=m_{\bar{p}}-B_{\bar{p}}$, and $B_{N}$ ($B_{\bar{p}}$) is the nucleon
($\bar{p}$) binding energy.
In the two-body c.m. frame $\vec{p}_N + \vec{p}_{\bar{p}} = 0$ and Eq.~\eqref{s} reduces to
\begin{equation} \label{Eq.:M}
\sqrt{s}=~m_{\bar{p}}+m_{N}-B_{\bar{p}}-B_{N}.
\end{equation}
This form of $\sqrt{s}$ was considered in Ref. \cite{Mishustin}. However, when the annihilation of the antiproton with a nucleon takes place in a nucleus, the momentum dependent term in Eq.~\eqref{s}
is no longer negligible~\cite{s} and provides additional downward energy shift to that stemming from the binding energies $B_{\bar{p}}$ and
$B_{N}$.
Taking into account averaging over the angles $(\vec{p}_N + \vec{p}_{\bar{p}})^2 \approx \vec{p}_N^{~2}+\vec{p}_{\bar{p}}^{~2}$,
Eq.~\eqref{s} can be rewritten as \begin{equation} \label{Eq.:J}
 \sqrt{s}= E_{th} \left(\!1-\frac{2(B_{\bar{p}} + B_{Nav})}{E_{th}} + \frac{(B_{\bar{p}}+ B_{Nav})^2}{E_{th}^2} - \frac{1}{E_{th}}T_{\bar{p}} - \frac{1}{E_{th}}T_{Nav} \!\right)^{1/2},
\end{equation}
where $E_{th}= 2m_N$, $B_{Nav}$ and $T_{Nav}$ is the average binding and average kinetic energy per nucleon, respectively,
and $T_{\bar{p}}$ represents the $\bar{p}$ kinetic energy. The kinetic energies of the nucleon and the antiproton were calculated as
the expectation values of the kinetic energy operator ${T}_j=-\frac{\hbar^2}{2 m_j^*} \triangle$, where $m^*_j=m_j-S_j$ is the
(anti)nucleon reduced mass.

In the studies of $K^-$-nuclear potentials \cite{s, kaony}, the momentum dependence in $\sqrt{s}$ was transformed into the density dependence. The nucleon kinetic energy was approximated within the Fermi gas model by $T_N(\frac{\rho_N}{\rho_0})^{2/3}$, where $T_N=23$~MeV,
and the kaon kinetic energy was expressed within the local density approximation by $T_K \approx -B_K-{\rm Re}{\cal V}_K(r)$, where ${\cal V}_K = V_K + V_{\rm C}$ and $V_{\rm C}$ is the $K^-$ Coulomb potential, which led
to the expression
\begin{equation} \label{Eq.:K}
 \sqrt{s}= m_N + m_K - B_{Nav}- \xi_N B_K + \xi_K {\rm Re}{\cal V}_K(r) - \xi_N T_N(\frac{\rho_N}{\rho_0})^{2/3},
\end{equation}
where $\xi_{N(K)}=\frac{m_{N(K)}}{m_N+m_K}$.
In more recent calculations of kaonic atoms \cite{kaonic atoms1, kaonic atoms2} and $\eta$-nuclear bound states \cite{etaS1, etaS2}, $\delta\sqrt{s} = \sqrt{s} - m_N - m_H $ was adjusted to respect the
low density limit $ \delta\sqrt{s} \rightarrow 0$ upon $\rho \rightarrow 0$:
\begin{equation} \label{Eq.:E}
 \delta\sqrt{s}= - B_{Nav}\frac{\rho_N}{\bar{\rho}_N}- \xi_N B_{H}\frac{\rho_N}{\rho_0} - \xi_N T_N(\frac{\rho_N}{\rho_0})^{2/3} + \xi_{H}{\rm Re}V_{H}(r) - \xi_N V_{\rm C}(\frac{\rho_N}{\rho_0})^{1/3},
\end{equation}
where $\bar{\rho}_N$ is the average nucleon density and  $H=K,\eta$ (for $\eta$ mesons, the last term in Eq.~(21) is zero).

The absorptive $\bar{p}$ potential used fully self-consistently in our calculations of $\bar{p}$-nucleus states
acquires the form
\begin{equation}
 {\rm Im}V_{\bar{p}} (r,\sqrt{s},\rho)=\sum_{\text{channel}} B_c f_{\text{s}}(\sqrt{s}) {\rm Im}V_{\text{opt}}(r).
\end{equation}

\section{Results}

We adopted the formalism introduced in Section \ref{model} to detailed calculations of ${\bar p}$ bound states in selected nuclei across the periodic table.
First, we did not consider the $\bar{p}$ absorption and explored various dynamical effects in these nuclei caused by the antiproton in the $1s$ nuclear
state using the G-parity motivated $\bar{p}$-meson coupling constants scaled by the factor $\xi$ (Eq.~(13)). We studied model dependence of the calculations, as well
as the effect of the ${\bar p}$ self-interaction. We confirmed previous findings of Mishustin et al. \cite{Mishustin} who had revealed that the insertion of the $\bar{p}$
into the nucleus causes significant polarization of the nuclear core.
Then, we took into account the ${\bar p}$ absorption in the nuclear medium and performed first fully self-consistent calculations of ${\bar p}$ nuclei using an optical
potential consistent with ${\bar p}$ atom data \cite{mares}. Selected results of our calculations are presented in the following subsections.

\subsection{Dynamical effects and model dependence}
In order to explore the extent of dynamical effects in the nuclear core due to the presence of $\bar{p}$, we performed \textit{static} as well as \textit{dynamical}
calculations of ${\bar p}$ nuclei.  In static calculations the antiproton source terms are omitted in the right hand sides of \emph{all} equations of motion \eqref{meson eq}. When the exact G-parity symmetry is assumed for the $\bar{p}$ coupling constants, the antiproton potential is about $800$~MeV deep in $^{16}$O calculated statically within the TM2 model. When calculated dynamically, the $\bar{p}$ potential reaches nearly $1700$~MeV in all nuclei considered. The dynamical effects are thus
considerable and should not be neglected. The binding energies of $\bar{p}$ in the $1s$ state are 1212.4~MeV in $^{16}$O$_{\bar{p}}$ (TM2 model) and 1107.5~MeV in $^{208}$Pb$_{\bar{p}}$ (TM1 model). The corresponding total binding energies are $B=1259.9$~MeV and $B=2651.2$~MeV for $^{16}$O$_{\bar{p}}$ and $^{208}$Pb$_{\bar{p}}$,
respectively (compare with $B=128.9$~MeV for $^{16}$O and $B=1634.8$~MeV for $^{208}$Pb). \\
The antiproton embedded in the nucleus causes its compression and the nuclear core density increases, particularly in the vicinity of $\bar{p}$ where
it reaches $\sim$~3--4 times the normal nuclear density. The effect is more pronounced in lighter nuclei where the antiproton, which is localized in the central region
of the nucleus up to $\approx 1.5$~fm, affects the whole nucleus. In heavier nuclei, the increase in the core density distribution is significant only in the central region of the nucleus, $r \leq 2$~fm.\\
\begin{figure}[t]
\begin{center}
\includegraphics[width=0.8\textwidth]{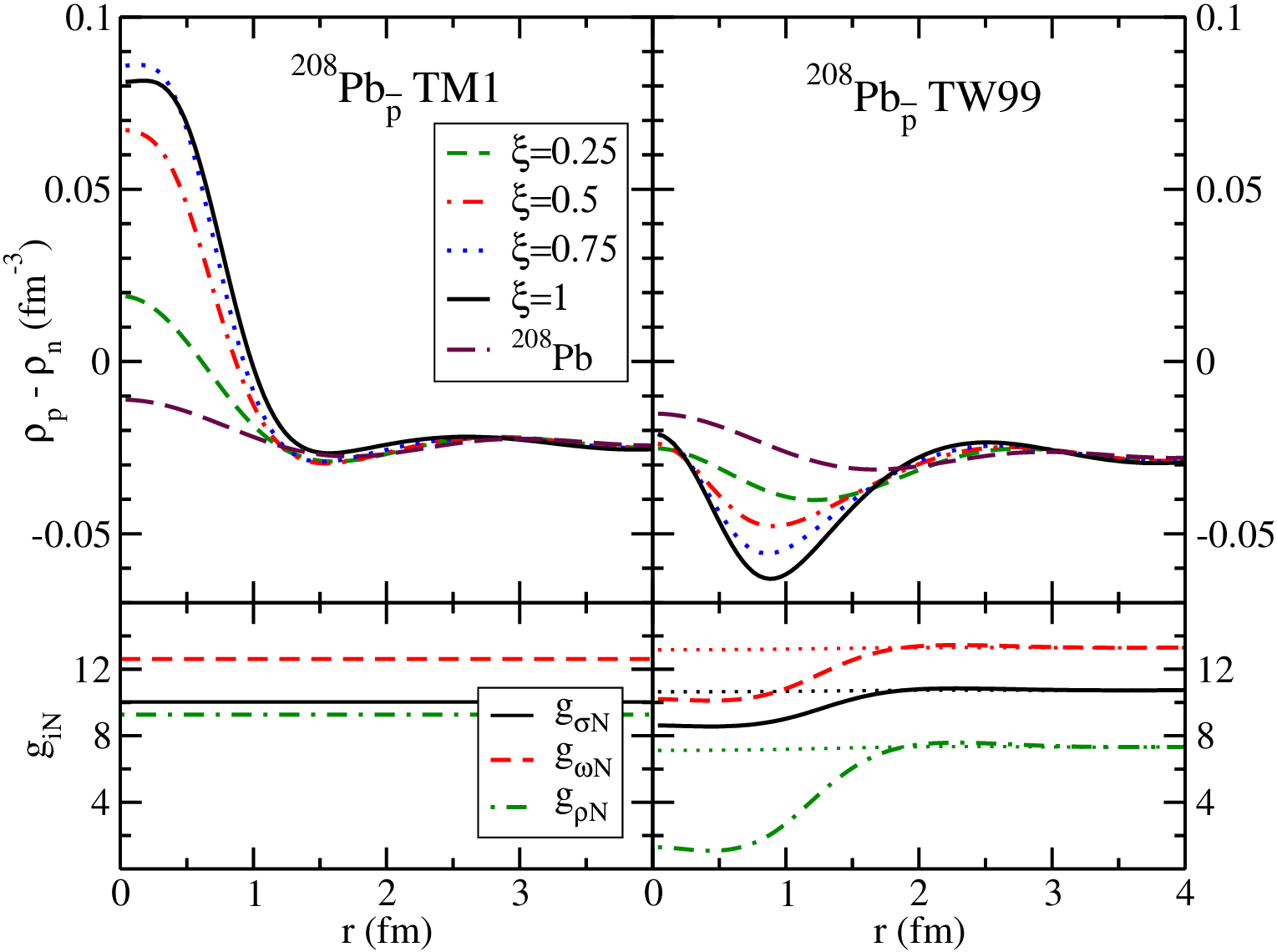}
\caption{\label{Fig.:DDmodel} The isovector density distribution in $^{208}$Pb$_{\bar{p}}$ calculated dynamically for different values of the scaling parameter $\xi$ within the TM1 model (upper left panel) and DD TW99 model (upper right panel). The radial dependence of the nucleon-meson couplings is shown in the lower panels. }
\end{center}
\end{figure}
Standard RMF models need not be reliable at such high nuclear densities occurring in ${\bar p}$ nuclei. Therefore, we also performed calculations using the density-dependent model TW99 \cite{TypelWolter} which is considered more suitable for the description of
dense nuclear matter. The TW99 model yields approximately the same depth of the $\bar{p}$ potential and somewhat higher nuclear core
densities than the TM model. This is due to the lower compressibility of the TW99 model -- compare $K=240$~MeV in the TW99 model,
$K=344$~MeV in the TM2 model, and $K=280$~MeV in the TM1 model. The only qualitative difference between the TM and TW99 models concerns the
isovector density distribution, $\rho_p(r) - \rho_n(r)$. In Fig. \ref{Fig.:DDmodel},  we present comparison of isovector densities in
$^{208}$Pb$_{\bar{p}}$ calculated dynamically within the TM1 and TW99 models for different values of the scaling factor $\xi$.
In the TM1 model (left panel), the density of protons exceeds the density of neutrons in the central region of the nucleus. Protons are more concentrated
around the $\bar{p}$ than neutrons because they feel strong isovector attraction which together with Coulomb attraction from the antiproton
surpass the Coulomb repulsion among protons. The rearrangement of the nuclear structure is sizeable even in light nuclei. In the TW99 model (right panel) we observe the
opposite effect. Neutrons are more concentrated in the center of the nucleus where the antiproton is localized. This is due to the decreasing
strength of the isovector $\rho$ meson coupling with increasing nucleon density as can be seen in the lower panel of~Fig.~\ref{Fig.:DDmodel}.
Consequently, protons feel much weaker isovector attraction and neutrons much weaker isovector repulsion in the center of
the nucleus. The isovector rearrangement of the nuclear structure is now less pronounced in light nuclei containing less nucleons.

\begin{figure}[t]
\begin{minipage}{0.47\textwidth} \vspace{-15pt}
\includegraphics[width=0.95\textwidth]{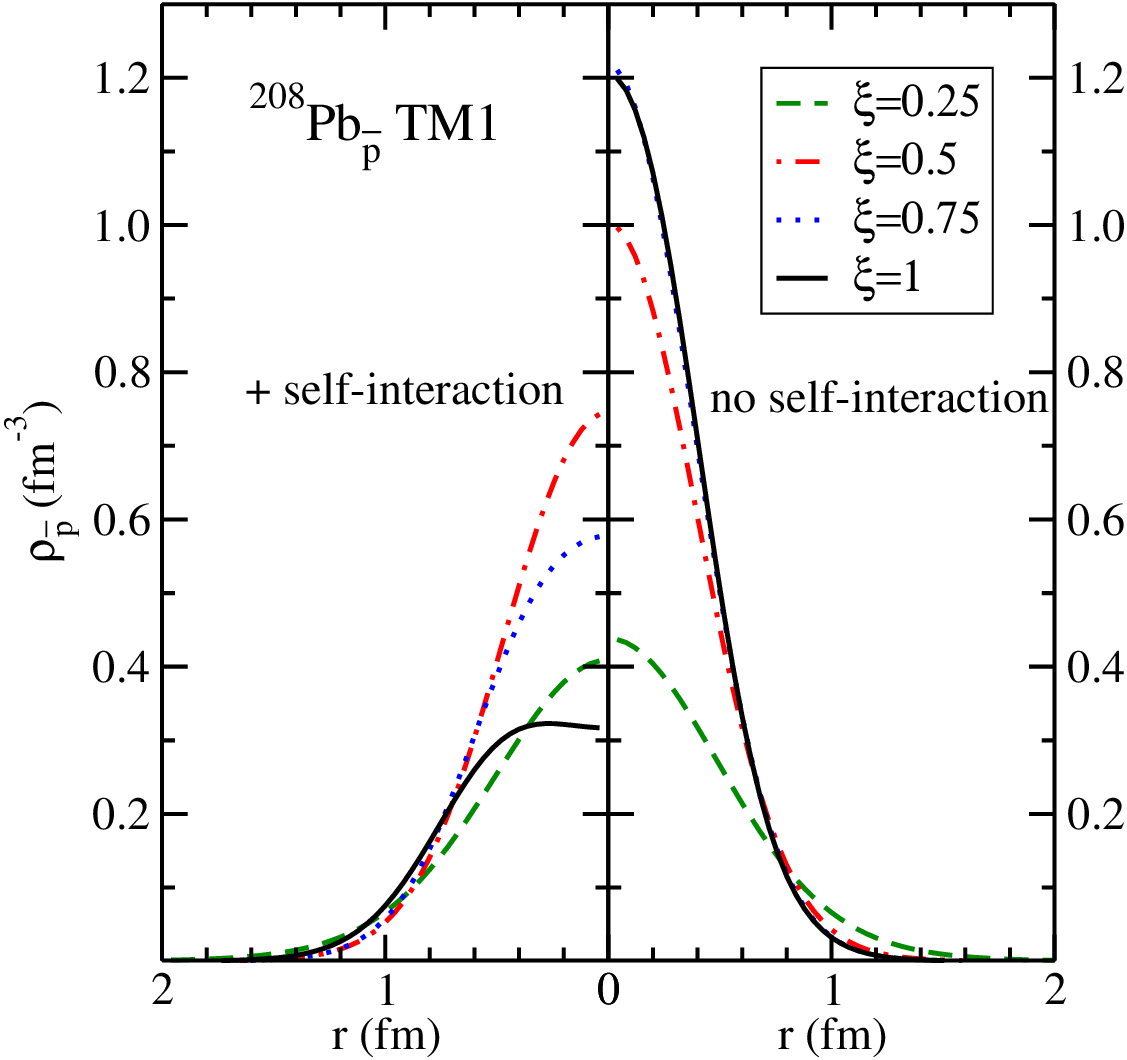}
\caption{\label{Fig.:rhoPbarSI} The $\bar{p}$ density distribution in $^{208}$Pb$_{\bar{p}}$, calculated dynamically for different values of $\xi$
in the TM1 model with (left) and without (right) the $\bar{p}$ self-interaction.}
\end{minipage}  \hspace{1.0pc}%
\begin{minipage}{0.47\textwidth}
\includegraphics[height=0.89\textwidth]{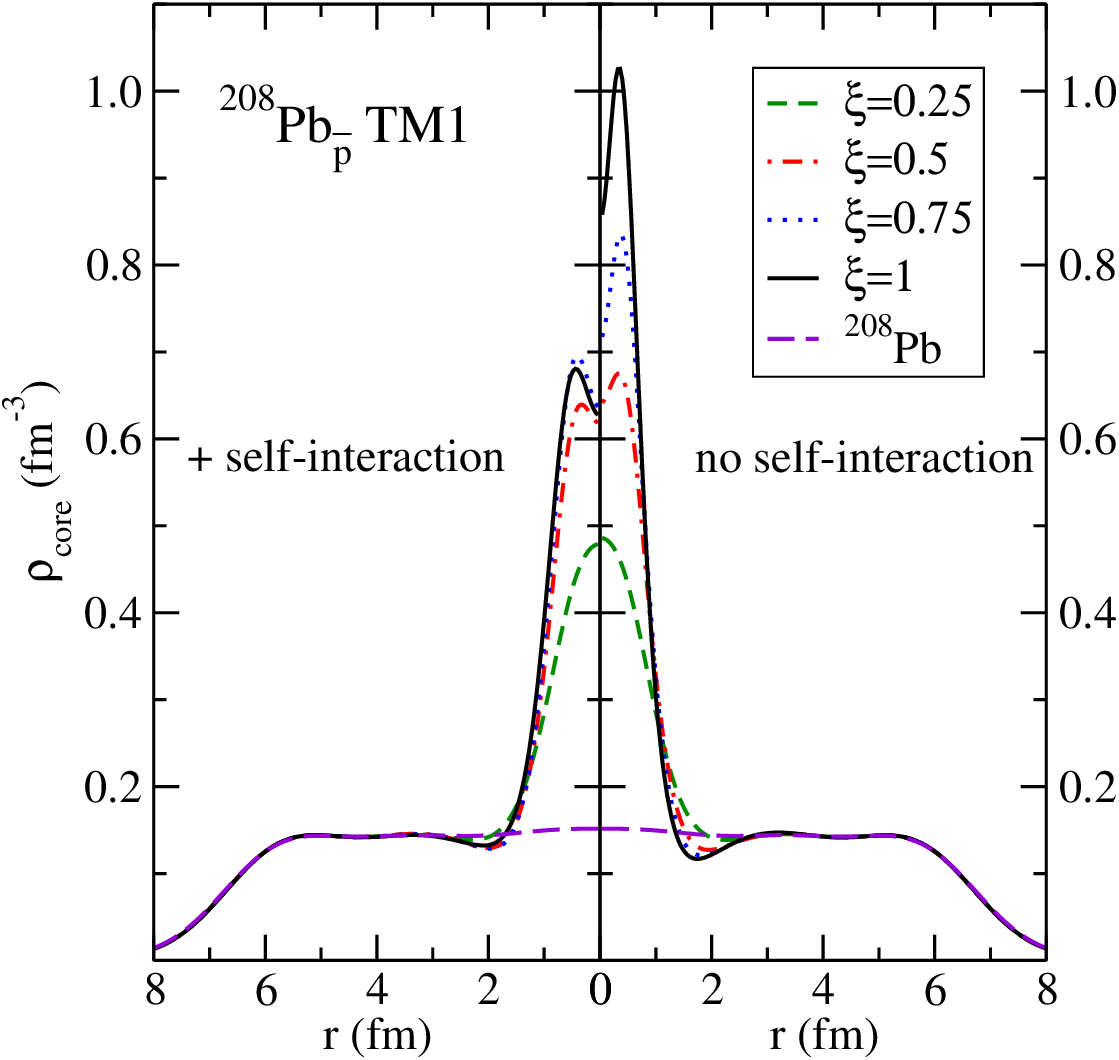}
\caption{\label{Fig.:rhoCoreSI}The core density distribution in $^{208}$Pb$_{\bar{p}}$, calculated dynamically for different values of $\xi$ in the
TM1 model with (left) and without (right) the $\bar{p}$ self-interaction. The case of $^{208}$Pb is shown for comparison.}
\end{minipage}
\end{figure}

During our dynamical calculations we noticed that the central ${\bar p}$ density $\rho_{\bar{p}}(0)$ reaches its maximum for $\xi \approx 0.5$ and then starts to decrease, as illustrated in the left part of Fig.~\ref{Fig.:rhoPbarSI}.
Here, we present the ${\bar p}$ density distribution in $^{208}$Pb$_{\bar{p}}$, calculated dynamically for different values of $\xi$ using the TM1 model.
This sudden decrease of the central ${\bar p}$ density is due to the $\bar{p}$ self-interaction  (see Section 2) which causes sizable effects on the
calculated observables when the ${\bar p}$ potential in the nuclear medium is very deep.
When the $\bar{p}$ self-interaction is subtracted (right panel) the $\rho_{\bar{p}}(0)$ increases gradually with $\xi$ and saturates at much higher values of  $\xi$.
Fig. \ref{Fig.:rhoCoreSI} shows the nuclear core density distribution in $^{208}$Pb$_{\bar{p}}$, calculated dynamically in the TM1 model with (left) and without (right)
the $\bar{p}$ self-interaction. It follows a similar trend as the $\bar{p}$ density distribution, but it saturates at a different value of $\xi$.

\begin{figure}[b!]
\begin{center}
\includegraphics[width=0.45\textwidth]{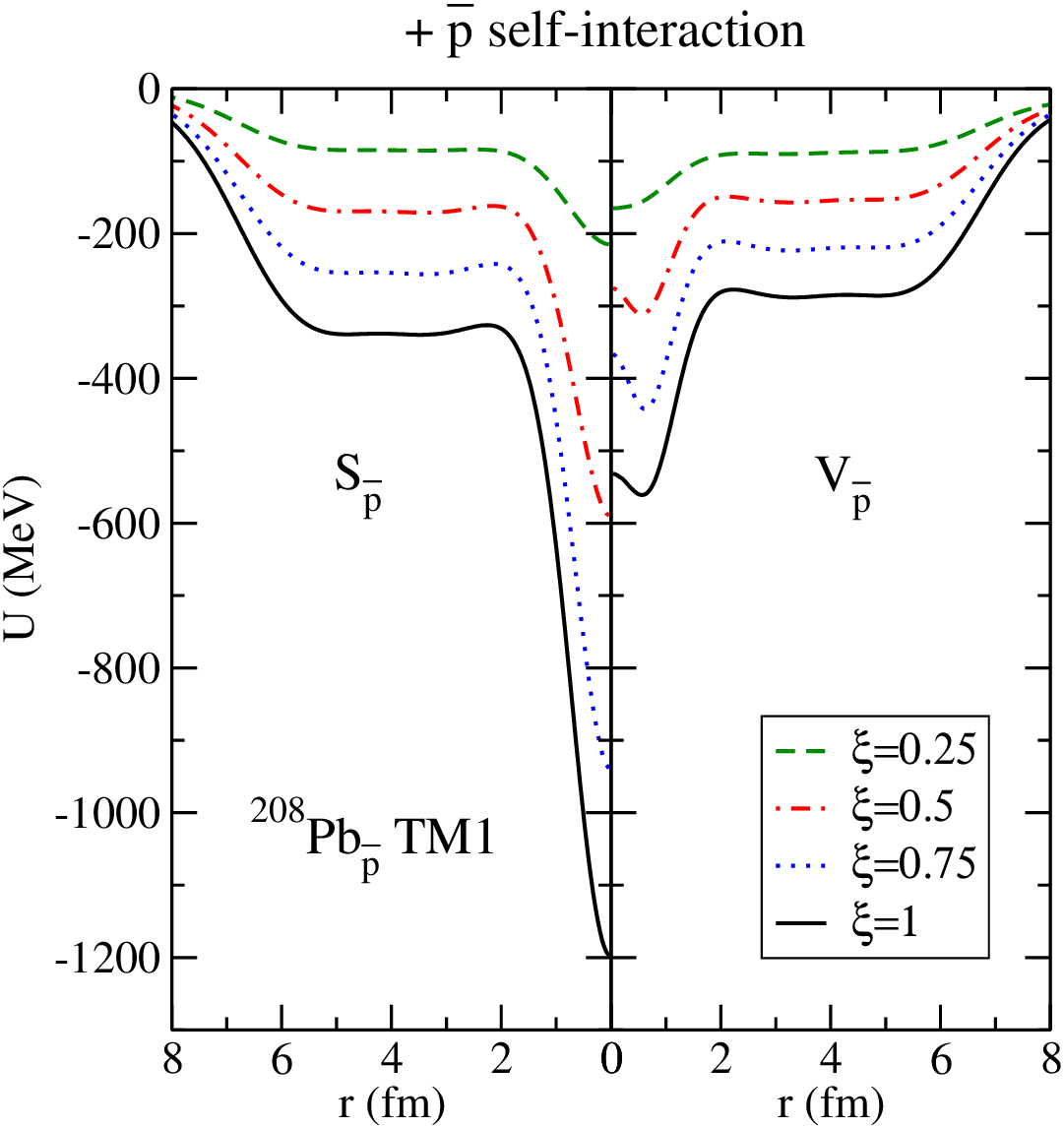}
\hspace{20pt}
\includegraphics[width=0.45\textwidth]{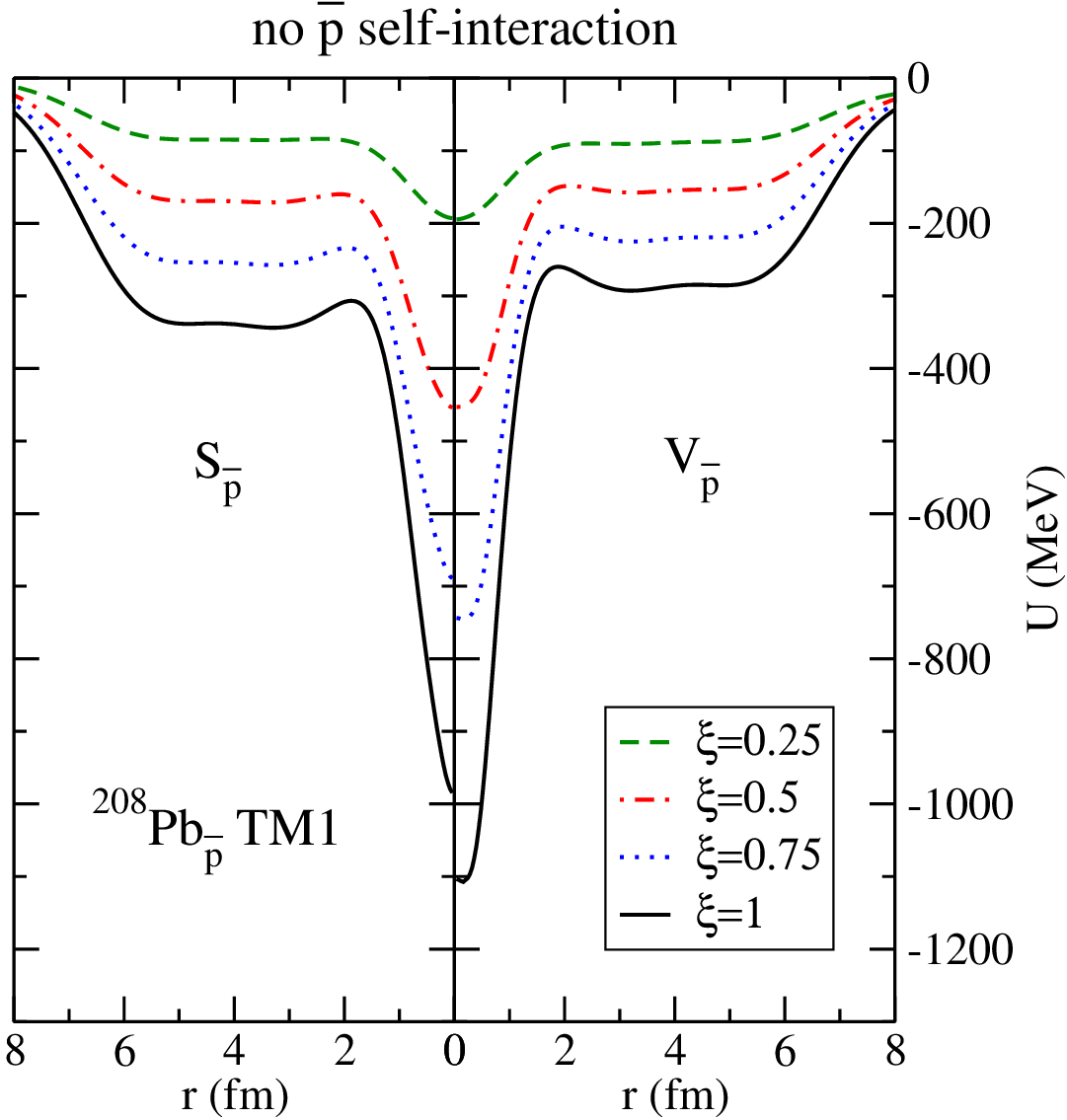}
\caption{\label{Fig.:SVpotSI} Scalar $S_{\bar p}$ and vector $V_{\bar p}$ potentials felt by $\bar{p}$ in $^{208}$Pb, calculated dynamically for different values of $\xi$ in the TM1 model with (left panel) and without (right panel) the $\bar{p}$ self-interaction.}
\end{center}
\end{figure}

In Fig.~\ref{Fig.:SVpotSI}, we compare scalar $S_{\bar p}$ and vector $V_{\bar p}$ potentials acting on the $\bar{p}$ in  $^{208}$Pb, calculated dynamically using the
TM1 model with and without the $\bar{p}$ self-interaction. When the $\bar{p}$ self-interaction is included (left panel),
the scalar potential $S_{\bar p}$  is deeper than the vector potential $V_{\bar p}$. The difference between their depths grows with increasing value of the scaling factor
-- for $\xi=1$, $S_{\bar p}(0)$ is twice as deep as $V_{\bar p}(0)$.
When the $\bar{p}$ self-interaction is subtracted (right panel), the $\bar{p}$ scalar potential is
comparable or even shallower than the vector potential, the difference between their depths being much smaller now.

\begin{figure}[tbh]
\begin{center}
\includegraphics[width=0.8\textwidth]{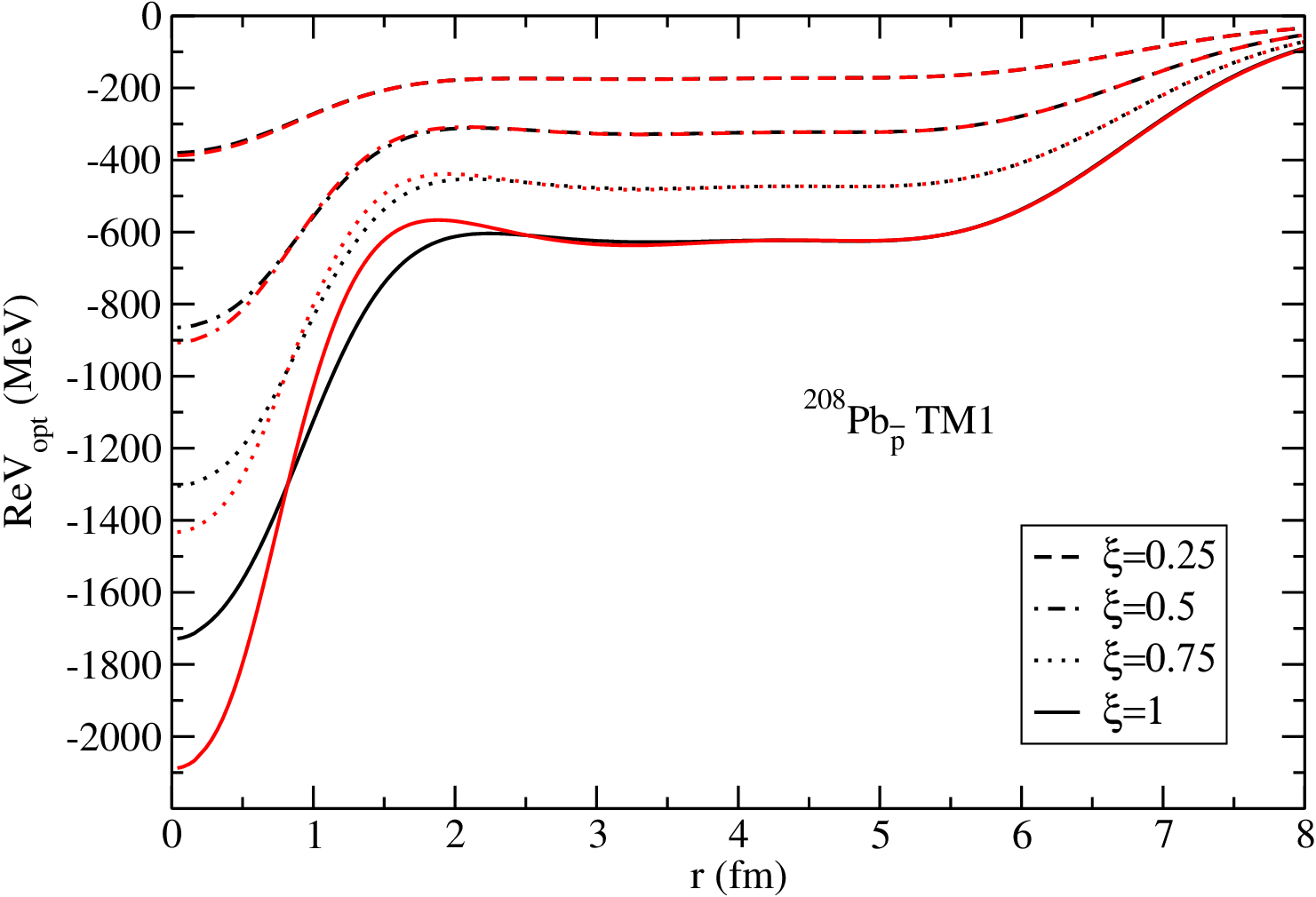}
\caption{\label{Fig.:S+V} The real part of the optical potential acting on ${\bar p}$ in
$^{208}$Pb, calculated dynamically for different values of $\xi$ in the TM1 model with (black lines) and without (red lines) the ${\bar p}$ self-interaction.}
\end{center}
\end{figure}

The interplay between the value of $\text{S}_{\bar{p}} - \text{V}_{\bar{p}}$,
the $\bar{p}$ single particle energy, and the $\bar{p}$ rest mass affects the large component of the solution of the Dirac equation for the underlying $\bar{p}$ wave function
which controls the density distribution. As the difference between the scalar and vector potential increases with $\xi$ in the case with the ${\bar p}$ self-interaction, a sudden change of sign occurs in the solution of the Dirac equation for the large component of the $\bar{p}$ wave function. Consequently, the density starts to decrease. It should be noted that the change of sign appears also in the case without the $\bar{p}$ self-interaction but at much higher values of $\xi$.

Fig.~\ref{Fig.:S+V} shows the total potential acting on ${\bar p}$ in $^{208}$Pb, calculated dynamically for selected values of $\xi$ within the TM1 model with (black lines) and without
(red lines) the ${\bar p}$ self-interaction. The effect of the $\bar{p}$ self-interaction starts to be considerable for really
deep $\bar{p}$ potentials, i.e. for $\xi \geq 0.5$. Correspondingly, the $\bar{p}$ binding energies and the total binding energies of $\bar{p}$ nuclei are
larger when the $\bar{p}$ self-interaction is subtracted and the effect increases with $\xi$ -- for $\xi=1$ the difference is more than $200$~MeV in
$^{208}$Pb$_{\bar{p}}$.

It is to be stressed that the available experimental data constrain the depth of the $\bar{p}$ potential at much lower values than the G-parity transformation.
The corresponding scaling factor of the $\bar{p}$ coupling constants which gives the potential consistent with the data is $\xi \approx 0.2$, which is safely in the region
where the effect of the ${\bar p}$ self-interaction is negligible.
From now on we will discuss the results of our calculations for the value of ${\xi} = 0.2$ only.

\begin{figure}[t!]
\begin{center}
\includegraphics[width=0.8\textwidth]{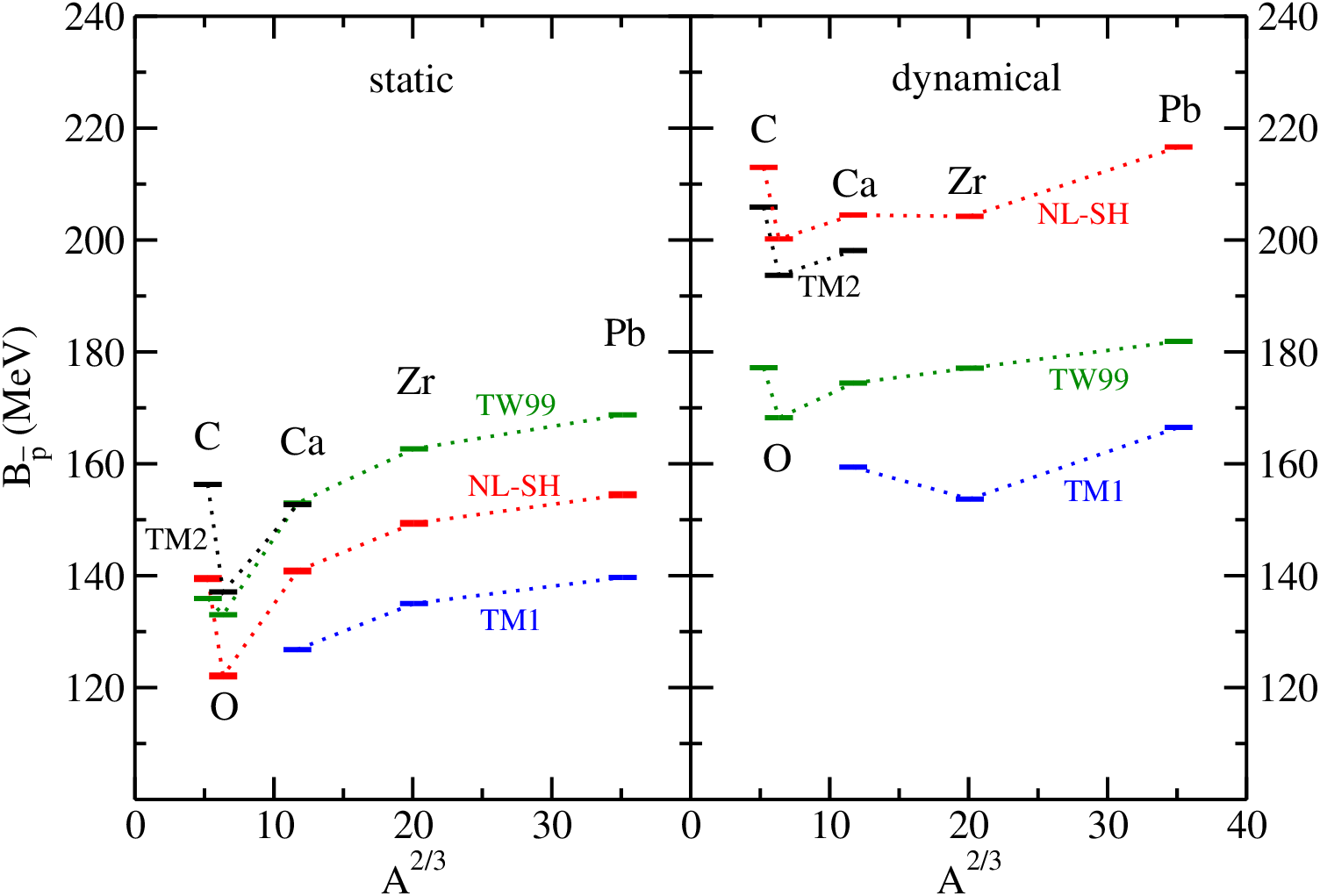}
\caption{\label{Fig.:8}Binding energies of $1s$ $\bar{p}$-nuclear states accross the periodic table calculated statically (left) and dynamically (right) using
the TM2 (black), TM1 (blue), NL-SH (red) and TW99 (green) models.}
\end{center}
\end{figure}

Binding energies $B_{\bar p}$ of $1s$ $\bar{p}$-nuclear states in core nuclei from $^{12}$C to $^{208}$Pb are plotted in Fig. \ref{Fig.:8}, where the results of static as well as dynamical calculations for various RMF models are presented.
Substantial differences between the ${\bar p}$ binding energies calculated statically and dynamically indicate
that the polarization of the nuclear core is, even for $\xi =0.2$, still significant.
Indeed, the central nuclear core densities are almost twice larger than the saturation density.
The $\bar{p}$ binding energies shown in the figure were calculated using the TM1, TM2, NL-SH and TW99 models. They
evince a strong model dependence.
In this work we often used the TM model \cite{Toki} which consists of two parameter sets -- the TM2 model designed to
account for properties of light nuclei and the TM1 model describing heavy nuclei. However, these two TM parametrizations yield quite
different characteristics of ${\bar p}$ nuclei, as illustrated in the figure.
There is a large inconsistency between $B_{\bar p}$ in light nuclei calculated using the TM2 model and $B_{{\bar p}}$ for
the TM1 model in heavy nuclei (compare also $B_{\bar p}$ in Ca for both TM1 and TM2).
In the case of the NL-SH and TW99 models the  $\bar{p}$  binding energy grows with increasing $A$, as expected, since the
antiproton feels attraction from larger amount of nucleons (except $^{12}$C with an extreme central density).
The differences between the ${\bar p}$ binding energies calculated statically and dynamically
indicate that the response of the nuclear core to the extra antiproton varies with the applied RMF model, where
nuclear compressibility seems to be the decisive factor.
The TW99 model gives the lowest value of the nuclear compressibility ($K=240$~MeV) out of the models used in our calculations. Consequently, there is a smallest difference between $B_{\bar p}$  calculated statically and dynamically. Then follow the TM1 and TM2
models with compressibilities $K=280$~MeV and $K=344$~MeV, respectively. The largest dynamical change of the $\bar{p}$ binding energy is observed for the NL-SH model with $K=355$~MeV.\\
As demonstrated in  Fig. \ref{Fig.:8}, the $\bar{p}$ binding energies calculated using the above RMF models remain sizable even for
the reduced $\bar{p}$ couplings ($\xi=0.2$), which has consequences for the evaluation of the widths of ${\bar p}$-nuclear states
discussed in the following subsection.

\subsection{$\bar{p}$ annihilation in a nucleus}

 We performed first fully self-consistent calculations of ${\bar p}$-nuclear states including antiproton absorption in a nucleus.
The $\bar{p}$ annihilation was described by the imaginary part of a phenomenological optical potential, parameters of which were
determined from global fits to antiproton atom data \cite{mares}. The effective scattering length Im$b_0$ (see Eq.~(14)) accounts for the
$\bar{p}$ absorption at threshold. However, the energy available for ${\bar p}$ annihilation products in the medium is lowered for the deeply bound antiproton. As a consequence, many annihilation channels may be considerably suppressed, which could result in significantly reduced widths of the deeply bound ${\bar p}$-nuclear states \cite{Mishustin}.
We evaluated the phase space suppression factors for considered annihilation channels as described in Section~2. They are presented in
Fig. \ref{Fig.: SF} as a function of the center-of-mass energy. As $\sqrt{s}$ decreases many channels become suppressed or even closed,
especially channels with massive particles in the final state and multi-particle decay channels. The range of $\sqrt{s}$ relevant for
our calculations,  $\sqrt{s} \approx 1.55\; {\text -}\; 1.72$~GeV, is denoted by the shaded area in Fig. \ref{Fig.: SF}.
Unlike Ref.~\cite{Mishustin}, we considered also kaon annihilation channels in our calculations. However, their contribution to the total
$\bar{p}$ width was found negligible ($5$~MeV at most).

\begin{figure}[t!]
\begin{center}
\includegraphics[width=0.45\textwidth]{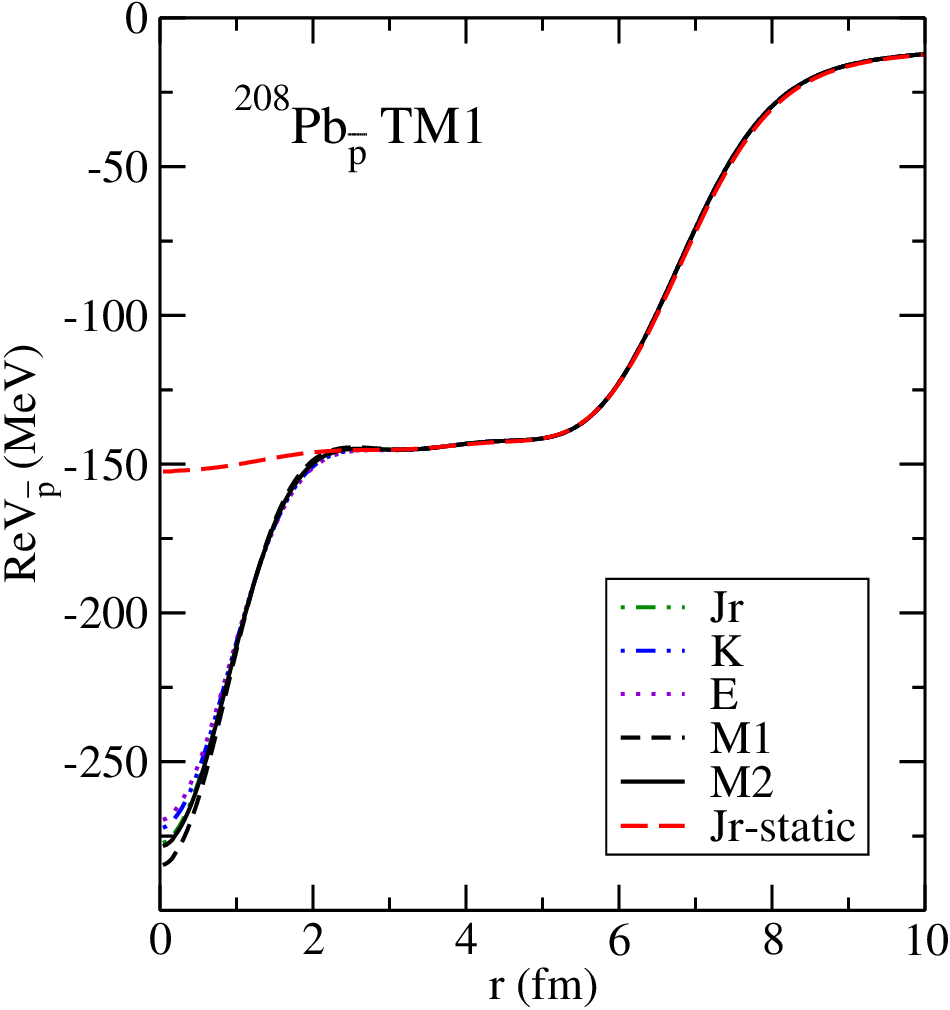}
\hspace{20pt}
\includegraphics[width=0.45\textwidth]{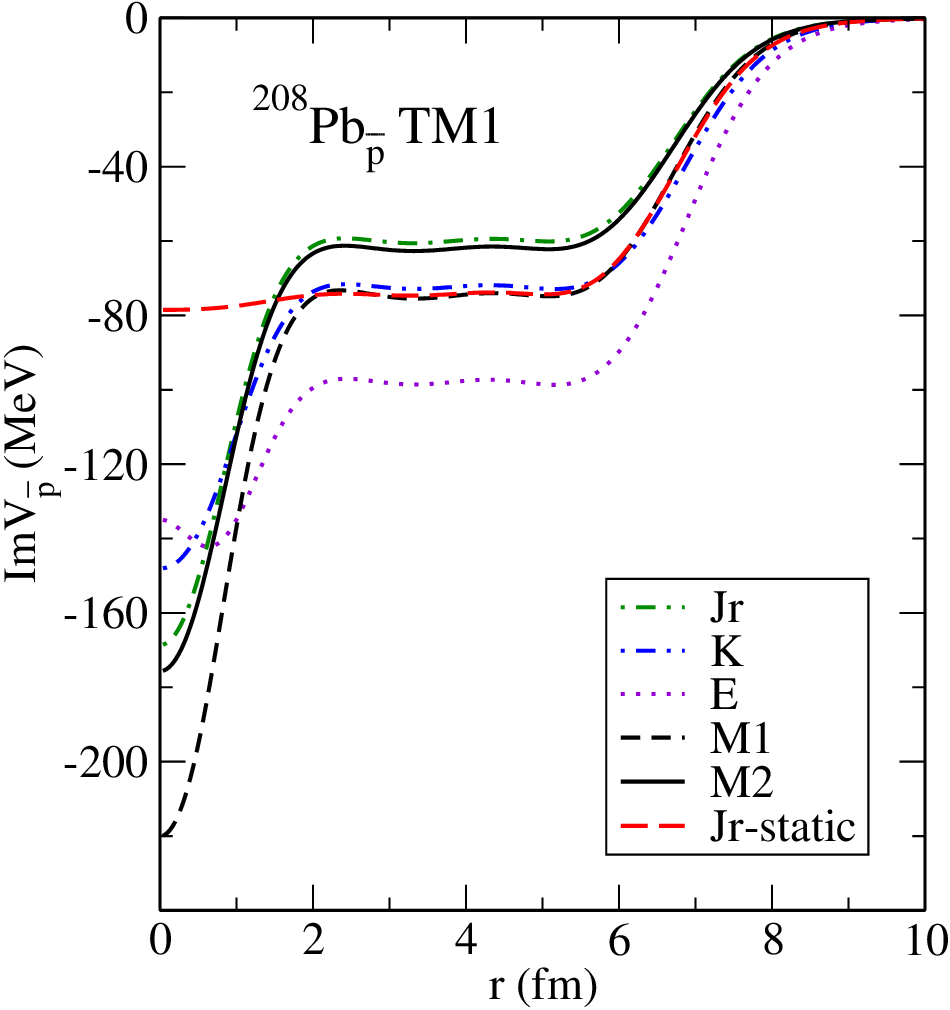}
\caption{\label{Fig.:5} The real Re$V_{\rm opt}$ and imaginary Im$V_{\rm opt}$ part of the $\bar{p}$ potential in $^{208}$Pb,
calculated dynamically using the TM1 model and different procedures of handling self-consistently  $\sqrt{s}$ (see text for details).
The $\bar{p}$ potential calculated statically for selected $\sqrt{s}$ (Jr-static) is shown for comparison.}
\end{center}
\end{figure}
We considered various procedures for handling $\sqrt{s}$ which controls the phase space reduction and consequently the ${\bar p}$
widths. First, we adopted $\sqrt{s}$ defined by Eq. \eqref{Eq.:M} which was applied by Mishustin et al. \cite{Mishustin}. We also assumed two
scenarios -- the annihilation with a proton in the $1s$ state, $B_N=B_{p1s}$, (denoted by M2) and the case when $B_N$ was
replaced by the average binding energy per nucleon $B_{Nav}$ (denoted by M1). Next, we used $\sqrt{s}$ transformed into the antiproton--nucleus system \eqref{Eq.:J} with non-negligible contribution from kinetic energies of annihilating partners.
To explore the effect of the medium, we calculated the underlying kinetic energies for constant (Jc) as well as reduced (Jr)
(anti)nucleon masses. Finally, we applied the forms of $\sqrt{s}$ used in the calculations of kaonic nuclei \eqref{Eq.:K}, and
$\eta$ nuclei as well as kaonic atoms~\eqref{Eq.:E} (denoted by K and E, respectively).
In Fig. \ref{Fig.:5}, we present the real and imaginary parts of the $\bar{p}$ potential in $^{208}$Pb, calculated dynamically in
the TM1 model for the above forms of $\sqrt{s}$. The real parts of the $\bar{p}$ potentials calculated dynamically
for $\xi=0.2$ and Im$b_0=1.9$~fm  have approximately the same depth for all considered procedures for evaluating $\sqrt{s}$.
On the other hand, the absorptive $\bar{p}$ potentials  Im$V_{\rm opt}$ exhibit strong dependence on the applied form of $\sqrt{s}$.
The $\bar{p}$ potential calculated statically is shown in the figure as well.
Both Re$V_{\rm opt}$ and Im$V_{\rm opt}$ are much shallower than the dynamically calculated potentials in the central region of the nucleus which illustrates the importance of
a dynamical, self-consistent treatment during antiproton-nucleus bound states calculations using an optical potential
describing the $\bar{p}$ absorption.

\begin{figure}[t]
\begin{center}
\includegraphics[width=0.45\textwidth]{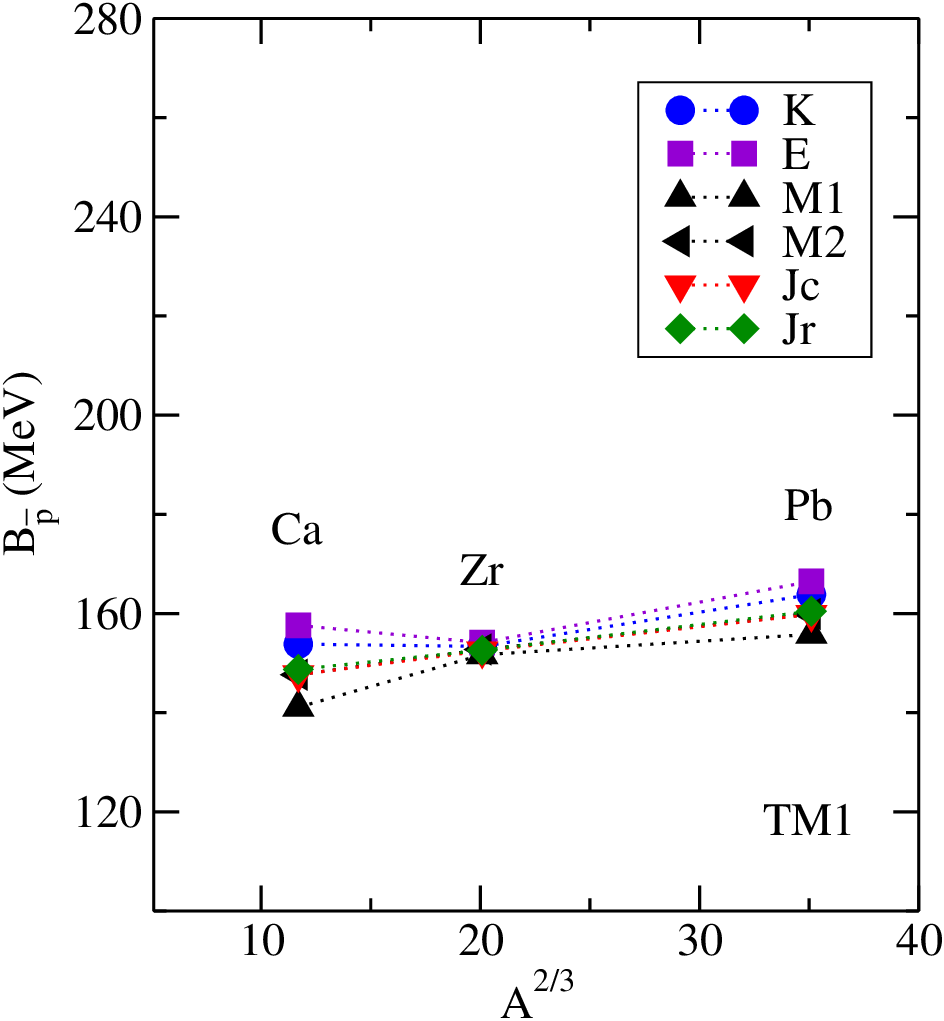}
\hspace{20pt}
\includegraphics[width=0.45\textwidth]{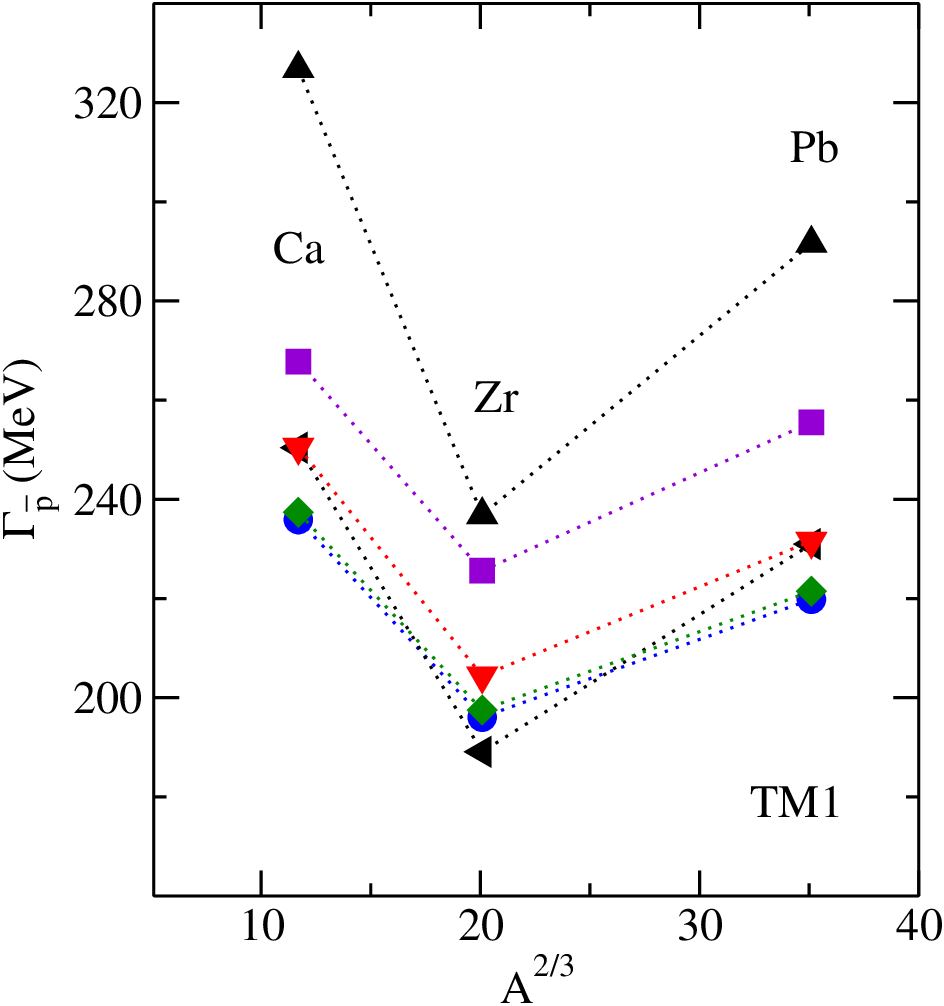}
\caption{\label{Fig.:65}Binding energies (left panel) and widths (right panel) of $1s$ $\bar{p}$-nuclear states in selected nuclei,
calculated dynamically using the TM1 model and different forms of $\sqrt{s}$ (see text for details).}
\end{center}
\end{figure}

In Fig. \ref{Fig.:65}, we compare binding energies (left panel) and widths (right panel) of $1s$ $\bar{p}$-nuclear states in
$^{40}$Ca, $^{90}$Zr, and $^{208}$Pb, calculated dynamically for $\xi=0.2$ and Im$b_0=1.9$~fm  using the same RMF model (TM1) but different forms of $\sqrt{s}$.
As can be seen, the $\bar{p}$ energies in a given nucleus calculated using different forms of $\sqrt{s}$ do not deviate much from
each other since the real parts of the underlying $\bar{p}$ potential are approximately the same (see Fig.~\ref{Fig.:5}, left panel).
The ${\bar p}$ widths are sizable and exhibit much larger dispersion. The largest widths are
predicted for $\sqrt{s}=$M1 and the corresponding $\bar{p}$ binding energies are thus the smallest. The ${\bar p}$ widths are
significantly reduced after including the non-negligible momentum dependent term in $\sqrt{s}$.
It is due to the additional sizable downward energy shift coming from the $\bar{p}$ and nucleon kinetic energies
\footnote{Similarly reduced $\bar{p}$ widths are obtained for $\sqrt{s}=$M2. However, in this case the annihilation of $\bar{p}$ with a proton in the $1s$ state is assumed ($B_{p1s} \gg B_{Nav}$).}.
The kinetic energies calculated with reduced masses ($\sqrt{s}=$Jr) are larger and consequently the $\bar{p}$ widths are smaller than those calculated using constant masses ($\sqrt{s}=$Jc); the difference is up to $15$~MeV in the TM1 model. The $\bar{p}$ widths calculated using $\sqrt{s}=$K and Jr are comparable.
However, when the low density limit is taken into account ($\sqrt{s}=$E) the $\bar{p}$ widths become by $\approx 30$~MeV larger.
\begin{figure}[t!]
\begin{center}
\includegraphics[width=0.45\textwidth]{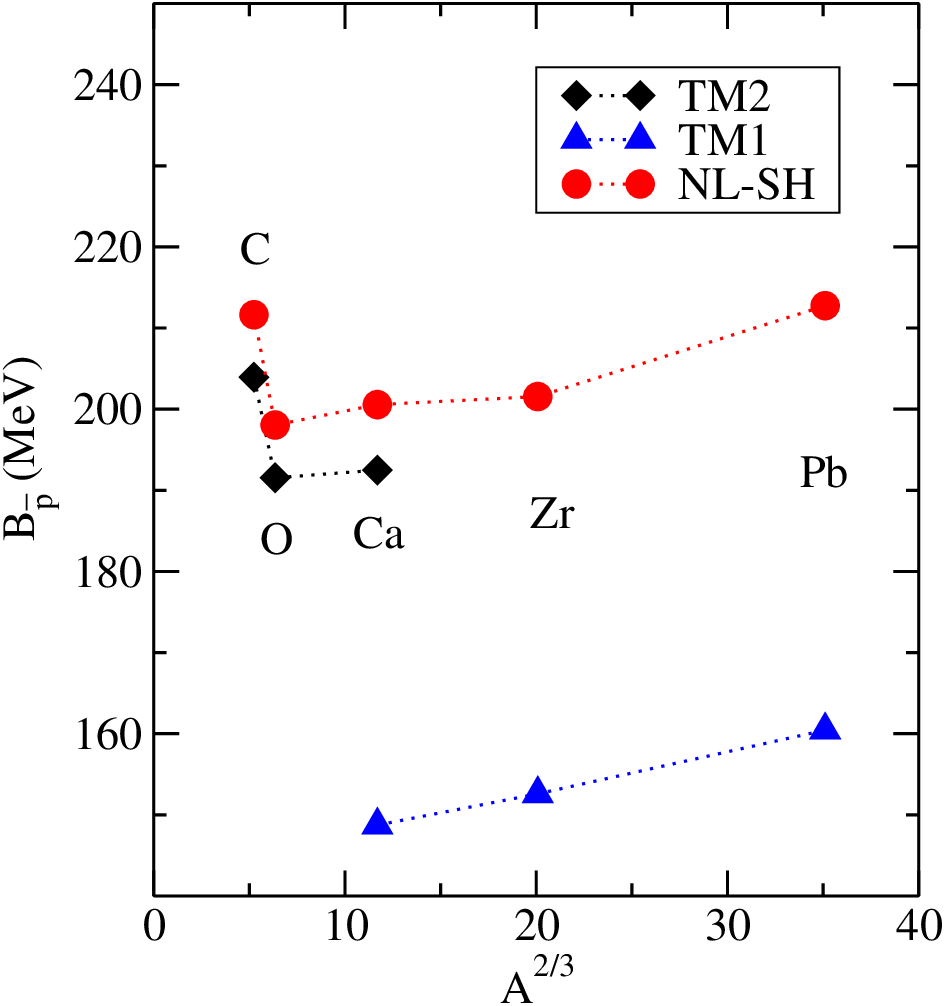}
\hspace{20pt}
\includegraphics[width=0.45\textwidth]{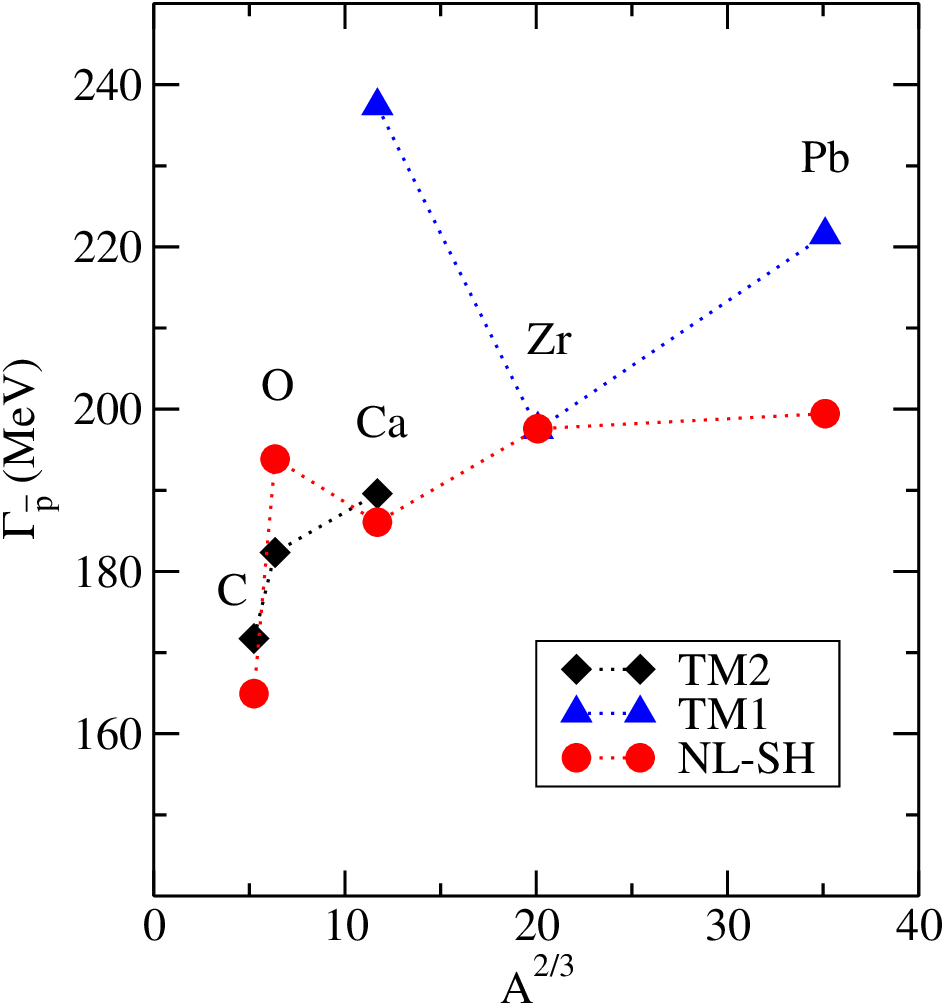}
\caption{\label{Fig.:66}Binding energies (left panel) and widths (right panel) of $1s$ $\bar{p}$-nuclear states across the periodic
table calculated dynamically for $\sqrt{s}=$Jr using the TM2 (black), TM1 (blue) and NL-SH (red) models.}
\end{center}
\end{figure}
\begin{center}
\begin{table}[b!]
\caption{\label{Tab.: results} Binding energies $B_{\bar{p}}$ and widths $\Gamma_{\bar{p}}$ (in MeV) of the $1s$ $\bar{p}$-nuclear
state in $^{16}$O, calculated dynamically (Dyn) and statically (Stat) within the TM2 model using the real and complex potentials consistent with $\bar{p}$--atom data (see text for details).}
\centering
\begin{tabular}{lcccccccc}
\hline\noalign{\smallskip}
 & \multicolumn{2}{c}{Real} & \multicolumn{2}{c}{Complex} & \multicolumn{2}{c}{$f_{\text{s}}({\rm M1})$} & \multicolumn{2}{c}
{$f_{\text{s}}({\rm Jr})$}  \\ \noalign{\smallskip}\hline\noalign{\smallskip}
 & Dyn & Stat & Dyn & Stat & Dyn &  Stat & Dyn & Stat  \\ \noalign{\smallskip}\hline\noalign{\smallskip}
$B_{\bar{p}}$ & ~193.7~ & ~137.1~ & ~175.6~ & ~134.6~ & ~190.2~ & ~136.1~ & ~191.5~ & ~136.3~ \\
$\Gamma_{\bar{p}}$ & ~-~ & - & ~552.3 & ~293.3 & ~232.5 & ~165.0 & ~182.3& ~147.0 \\
\noalign{\smallskip}\hline
\end{tabular}
\end{table}
\end{center}
The model dependence of the $\bar{p}$ binding energies and widths of $1s$ $\bar{p}$-nuclear states across the periodic
table calculated dynamically for $\xi=0.2$, Im$b_0=1.9$~fm, and  $\sqrt{s}=$Jr is illustrated in Fig. \ref{Fig.:66}.
The TM2 and NL-SH models give similar $\bar{p}$ binding energies in $^{12}$C, $^{16}$O and $^{40}$Ca. The corresponding ${\bar p}$
widths are also quite close to each other. On the other hand, the TM1 model, which yields considerably lower values of
$B_{\bar p}$ predicts larger ${\bar p}$ widths than the TM2 and NL-SH models (except the case of $^{90}$Zr).

In Table \ref{Tab.: results}, we present binding energies $B_{\bar{p}}$ and widths $\Gamma_{\bar{p}}$  of the $1s$ $\bar{p}$-nuclear
state in $^{16}$O, calculated using the real and complex potentials consistent with $\bar{p}$-atom data ($\xi=0.2$, Im$b_0=1.9$~fm).
To illustrate the role of the suppression factors $f_s$ we show the results of calculations without $f_s$ (`Complex'), as well as
including $f_s$ for $\sqrt{s}$ due to $B_{\bar{p}}$ and $B_{Nav}$ ('$f_s$(M1)') and for $\sqrt{s}$ with the additional downward
energy shift caused by the momenta of annihilating partners ('$f_s$(Jr)').
The static calculations, which do not account for the core polarization effects, give
approximately the same values of the $\bar{p}$ binding energy for all cases. The binding energies calculated
dynamically are much larger, which indicates that the polarization of the core nucleus is significant. When the phase space suppression
is taken into account the $\bar{p}$ width is reduced by more than twice (compare `Complex' and '$f_s$(M1)' in the last row of
Table~\ref{Tab.: results}).
When treating $\sqrt{s}$ self-consistently including the $\bar{p}$ and $N$ momenta (see '$f_s$(Jr)'), the $\bar{p}$ width is
reduced by additional $\approx50$~MeV, but still remains sizable.
The corresponding lifetime of the $\bar{p}$ in the nucleus is $\simeq 1$~fm/c.

\section{Conclusions}

In this work, we studied the sub-threshold antiproton interaction with the nuclear medium. The real part of the ${\bar p}$-nucleus potential
was constructed within the RMF approach using G parity as a starting point. Since the empirical ${\bar p}$-nucleus
interaction is much weaker than that derived from G-parity transformed ${\bar p}$ coupling constants, a uniform scaling factor $\xi $ was
introduced to control the strength of the $\bar p$-nucleus interaction.

We explored dynamical effects caused by the presence of the strongly interacting ${\bar p}$ in the $1s_{1/2}$ state of selected nuclei
across the periodic table and confirmed sizable changes in the nuclear structure. The central density of the nuclear core considerably
increases -- it reaches about 3 times the normal nuclear density. While in light nuclei the antiproton affects the entire nucleus, in
heavier nuclei the increase in the core density distribution is significant only in the central region where ${\bar p}$ is localized,
$r \leq 2$~fm. Since various RMF models give quite different equation of state at such high densities, we employed several RMF
parametrizations including the density-dependent TW99 model to check the model dependence of our results.
The response of the nuclear core to the strongly bound antiproton varies with the applied RMF model as it is affected by
the corresponding nuclear compressibility.

In the RMF approach, the antiproton as well as each nucleon moves in mean fields created by all (anti)nucleons in the
nucleus, including itself. The effect of the ${\bar p}$ self-interaction increases with the strength of the ${\bar p}$ couplings. It
causes saturation of the antiproton and nuclear core density distributions and subsequent decrease at some critical value of the
scaling factor $\xi$. We checked that for the values of $\xi\sim0.2\; \text{-}\; 0.3$, consistent with empirical $\bar{p}$-nucleus potentials with
depths $\approx 150\; {\text -}\; 200$~MeV, the effect is tiny and can thus be neglected. This finding is general enough to be applied in RMF calculations
of other nuclear systems with a strongly interacting hadron.

In order to include the $\bar{p}$ annihilation in the nuclear medium, we adopted the imaginary part of a phenomenological optical
potential with parameters constrained by fits to $\bar{p}$-atom data.  We considered various relevant decay
channels of $\bar{p}N$ annihilation at rest and took into account the phase space suppression for annihilation products of the deeply bound antiproton in the nuclear medium.
We performed dynamical calculations of ${\bar p}$-nuclear bound states using a complex optical potential consistent with ${\bar p}$-atom data.
We explored in detail the interplay between the underlying dynamical processes and the relevant kinematical conditions that determine the
annihilation width of ${\bar p}$ bound states in the nuclear medium.
The $\bar{p}$ widths decrease by factor 2 when the suppression of the phase space is considered and they are further reduced by
$\approx 50$~MeV when the momenta of annihilating partners are taken into account. However, the
 ${\bar p}$ widths still remain sizeable for a realistic $\bar{p}$-nucleus interaction. We noticed that the $\bar{p}$ absorption remarkably
influenced the polarization of the nuclear core. It is therefore mandatory to perform the calculations with a complex $\bar{p}$-nucleus
potential fully self-consistently. Such calculations were performed in this work for the first time ever.

It is desirable to use the self-consistent techniques applied in this work in calculations of ${\bar p}$-nucleus
interaction based on a more fundamental ${\bar N}N$ potential model, such as the Paris ${\bar N}N$ potential \cite{Paris2} used in the most
recent study of ${\bar p}$ atoms \cite{Paris1}, and compare them with the calculations within the RMF approach. We are currently finalizing  such
calculations and the results will be published elsewhere.
It is also desirable to study in detail the $\bar{p}$-nucleus interaction above threshold to describe
${\bar p}$-nucleus scattering processes because knowledge of such processes, of the ${\bar p}$ behavior in the nuclear medium, as well as
post ${\bar p}$ annihilation dynamics of the nuclear core is expected to be in great demand in view of future experiments at FAIR \cite{FAIR}. Considering anticipated production of hyperon-antihyperon pairs in $\bar{p}$-nucleus collisions at FAIR it is timely to extend the present model to calculations of nuclear systems with (anti)hyperons.

\section*{Acknowledgements}
This work was supported by the GACR grant No.~ P203/15/04301S. J.M. acknowledges financial support within the agreement on
scientific collaboration between the Czech Academy of Sciences
and the Israel Academy of Sciences and Humanities. Both J.H. and J.M. acknowledge the hospitality extended to them at the
Racah Institute of Physics, The Hebrew University of Jerusalem, during a collaboration visit in April/May 2015.
We wish to thank E. Friedman, A. Gal, and S. Wycech for valuable discussions, and P. Tlust\'{y} for his assistance during Monte
Carlo simulations using PLUTO.

\end{document}